\documentclass[journal]{IEEEtran}

\ifCLASSINFOpdf
\else
\fi
\usepackage{bbm}
\usepackage{amsmath}
\usepackage{mathrsfs}
\usepackage{graphicx}          % Include this line if your
\newtheorem{theorem}{Theorem}[section]
\newtheorem{definition}{Definition}[section]

\newtheorem{example}{Example}[section]

\newtheorem{lemma}{Lemma}[section]
\newtheorem{remark}{Remark}[section]

\usepackage{algorithm}
\usepackage{graphics} % for pdf, bitmapped graphics files
\usepackage{epsfig} % for postscript graphics files
\usepackage{mathptmx} % assumes new font selection scheme installed
\usepackage{times} % assumes new font selection scheme installed
\usepackage{amsmath} % assumes amsmath package installed
\usepackage{amssymb}  % assumes amsmath package installed
\usepackage{colortbl}
\usepackage{stfloats}
\usepackage{colortbl}
\usepackage{epsfig}
\usepackage{latexsym}
\usepackage{mathrsfs}
\usepackage{psfrag}
\usepackage{graphicx,subfigure}
\usepackage{graphicx}
\usepackage{cite}
\usepackage{colortbl}
\usepackage{booktabs}
\usepackage{float}
\usepackage{blkarray}
\usepackage{stfloats}
\usepackage{arydshln}
\usepackage{setspace}
\usepackage{extarrows}

\begin{document}
%\doublespacing
%\begin{spacing}{1.5}
\title{A New Approach to Pinning Control of Boolean Networks}

\author{Jie Zhong, Daniel W.C. Ho, \emph{Follow, IEEE} and Jianquan Lu$^\ast$, \emph{Senoir Member, IEEE}% <-this % stops a space
\thanks{This work was supported by Research Grant Council of HKSAR under Grant Nos. GRF CityU 11200717, and the National Natural Science Foundation of China under Grant No. 61903339 and 61573102, the Natural Science Foundation of Jiangsu Province of China under Grant BK20170019.}\\
\thanks{$^\ast$ Corresponding author: Jianquan Lu}
\thanks{Jie Zhong is with the College of Mathematics and Computer Science, Zhejiang Normal University, Jinhua 321004, China (e-mail: zhongjie0615@gmail.com)}
%\thanks{Bowen Li is with the School of Information Science and Engineering, Southeast University, Nanjing 210096, China, and also with the School of Cyber Science and Engineering, Southeast University, Nanjing 210096, China (e-mail:qfhxjy@126.com)}
\thanks{Daniel W.C. Ho is with the Department of Mathematics, City University of Hong Kong, Kowloon, Hong Kong (e-email: madaniel@cityu.edu.hk).}
\thanks{Jianquan Lu is with the School of Mathematics, Southeast University, Nanjing
	210096, China (e-mail: jqluma@seu.edu.cn)}
% <-this % stops a space
%\thanks{Weihua Gui is with School of Automation, Central South University, Changsha 410083, China (email: gwh@csu.edu.cn)}
}
% <-this % stops a space


\markboth{Journal of \LaTeX\ Class Files,~Vol.~14, No.~8, August~2015}%
{Shell \MakeLowercase{\textit{et al.}}: Bare Demo of IEEEtran.cls for IEEE Journals}



\maketitle

\begin{abstract}
Boolean networks (BNs) are discrete-time systems where nodes are inter-connected (here we call such connection rule among nodes as network structure), and the dynamics of each gene node is determined by logical functions. In this paper, we propose a new approach on pinning control design for global stabilization of BNs based on BNs' network structure, named as network-structure-based distributed pinning control. By deleting the minimum number of edges, the network structure becomes acyclic. Then, an efficient distributed pinning control is designed to achieve global stabilization. Compared with existing literature, the design of pinning control is not based on the state transition matrix of BNs. Hence, the computational complexity in this paper is reduced from $O(2^n\times 2^n)$ to $O(2\times 2^K)$, where $n$ is the number of nodes and $K\leq n$ is the largest number of in-neighbors of nodes. In addition, without using state transition matrix, global state information is no longer needed, the design of pinning control is just based on neighbors' local information, which is easier to be implemented. The proposed method is well demonstrated by several biological networks with different sizes. The results are shown to be simple and concise, while the traditional pinning control can not be applied for BNs with such a large dimension.
\end{abstract}


% \fi
\IEEEpeerreviewmaketitle

%\usepackage{setspace}%ʹÓüä¾àºê°ü
%\begin{spacing}{2.5}%ʹÓüä¾àºê°ü
\section{Introduction}
In genetic regulatory networks, the behavior of genes or cells generates a logical phenomenon of activities, like active/inactive, expressed/unexpressed. As a typical logical system, Boolean network (BN) was first proposed by Kauffman to discover the internal behaviors of genes in genetic regulatory networks \cite{KauffmanS1969Jotbp437}. Since then, investigations on Boolean networks (BNs) have been a hot-spot to explore evolution patterns and structures \cite{Gupta2007JTBp463}. In a BN each node representing a cell or a gene takes a value from two logical variables 1 and 0. In addition, the state evolution of each node is determined by a logical function described by logical operators and its neighboring nodes. This implies that the dynamical behavior of a BN is determined by both network structure describing the neighbor relationship between gene nodes and also by the value update logical functions. Since then, BNs have attracted great attention such as in the study of the oscillations of signal transduction networks \cite{Saadatpour2010JoTBp641} and identifying BNs' structure \cite{Melkman2017IToNNaLSp,Azuma2017ITCNSp179,Azuma2019ITCNSp464}. In addition, many fundamental results have been obtained to disclose the relationships between network structure of a BN and its fixed points and cyclic attractors \cite{Mori2017PRLp028301}.

Recently, a new matrix product called semi-tensor product (STP) approach was proposed in \cite{ChengD2010pX}, which brings a new direction for further study of BNs. The STP of matrices is an extension of the conventional matrix product, and breaks the traditional dimension matching condition of matrix product. It was firstly proposed by Cheng and his colleagues \cite{ChengD2010pX} as a convenient tool to analyze logical functions, BNs and other finite-valued systems. Based on a bijective equivalence between logical variables and Boolean vectors, any logical function can be expressed in algebraic form. Under the framework of algebraic forms, a great deal of results have been established in the past few decades \cite{Laschov2012Ap1218,chen2019ITCNSp1379,wu2017ITCNSp1100,Meng2017ITNNLSp}. Some fundamental problems concerning Boolean (control) networks have been widely addressed, including probabilistic Boolean networks \cite{ShmulevichI2002Bio18}, observability \cite{zhang2016ITCNSp2733}, stability \cite{ChengD2011IJoRaNCp134,LiR2013IToACp1853,GuoY2015Ap106,zhou2015Ap230}. 
%In addition, partially-observed Boolean systems have been paid much attention, which are used to model the stochasticity in both states and measurement processes \cite{Mcclenny2017BBp519,Imani2018Ap238,Imani2017ITSPp359,Imani2018Ap172}. 

In biological systems or genetic networks, it is important to design therapeutic interventions that steer patients to desirable states, such as a healthy one, and maintain this state afterward \cite{Zanudo2015PCBpe1004193}. For example, some kinds of control actions (node and edge deletions) have been proposed to identify control targets in some biological networks, such as p53-mdm2 network and T cell lymphocyte granular leukemia survival signaling network \cite{Murrugarra2016BSBp94}. In \cite{Murrugarra2016BSBp94}, node deletion can model the control action of knockout of a node, while edge deletion can model the action of a drug that inactivates the corresponding interaction among two gene products.

Over the past years, stability and stabilization of BNs have attracted much attention. For example, stabilization of Boolean control networks (BCNs) based on two kinds of controls (open-loop control and state feedback control) has been addressed in \cite{ChengD2011IJoRaNCp134}. Later, the results have been extended to the issue of stabilization to some periodic cycles \cite{FornasiniE2013Ap1506}. In \cite{LiR2013IToACp1853}, Li \emph{et al.} firstly proposed an approach to design state feedback stabilization for BCNs, and subsequently Li \emph{et al.} presented a new way to design all possible feedback stabilizers in \cite{Li2017Ap303}. However, in these related papers, controllers are imposed to all nodes or applied to some nodes of BCNs. Due to the introduction of pinning control technique \cite{LiF2015IToNNaLSp1585,LuJ2016IToACp1658}, stabilization of BCNs under pinning controllers has been a new research direction \cite{Li2017ITNNLS,Li2017ITCYT}. For example, in \cite{LiF2015IToNNaLSp1585}, stabilization under pinning controllers has been studied, and an algorithm to design pinning controllers is firstly presented based on the state transition matrix of BNs.

Given a BN with $n$ nodes, using STP method, one can obtain its algebraic form: $x(t+1)=Lx(t)$ (matrix $L$ is called the state transition matrix with dimension $2^n\times 2^n$). According to the traditional pinning control method based on the state transition matrix $L$ (named as $L$-based pinning control) proposed in \cite{LiF2015IToNNaLSp1585}, $L$-based pinning control can be designed to achieve stabilization by changing columns of matrix $L$. This implies that the information of matrix $L$ is always needed in \cite{LiF2015IToNNaLSp1585}, and one cannot design pinning controllers without matrix $L$. Most of the results concerning the design of pinning controllers are based on this method; see \cite{LiF2015IToNNaLSp1585,LiF2016IToCaSIEBp309,LuJ2016IToACp1658,Li2017ITNNLS,Li2017ITCYT,li2020ITCNSp201}. However, the dimension of matrix $L$ grows dramatically with the number of nodes. Thus, traditional $L$-based pinning control method seems to be hard for implementation on some large-scale genetic networks, which is one of the main drawbacks. 
In addition, since $L$-based pinning control method is based on the information of matrix $L$, the pinning controllers are determined based on global state information of all the nodes. Traditional $L$-based pinning control design is in the form of $u(t)=g(x_1(t), \dots, x_n(t))$, where $x_1(t), \dots, x_n(t)$ are global states of a BN. This is another drawback of traditional $L$-based pinning control design, which will lead to a high dimensional controller design. To the best of our knowledge, there is no result available for studying stabilization of BNs based on network structure of BNs and using neighbors' local information. 

\begin{figure}\centering
	\includegraphics[width=7.6cm,height=6cm]{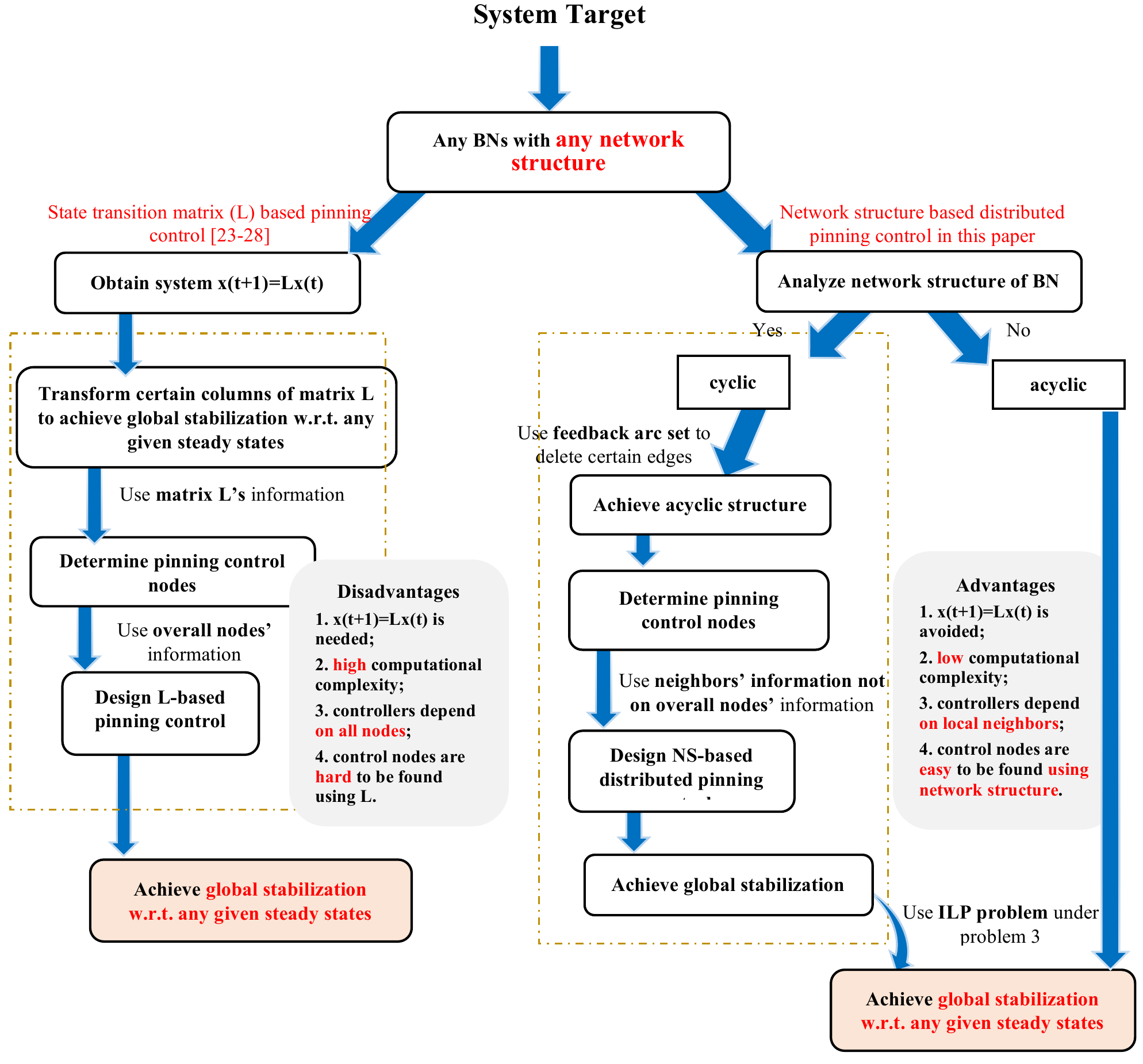}
	\caption{The summarized comparisons between NS-based distributed pinning control and traditional $L$-based pinning control.}\label{comparisons}
\end{figure}

In our previous paper \cite{LuJ2016IToACp1658}, the goal was to study the controllability problem of BCNs where some specific nodes are selected to be controlled. However, \cite{LuJ2016IToACp1658} does not present a method to determine the pinning controlled nodes and design the corresponding controllers. In addition, the controllability criteria in \cite{LuJ2016IToACp1658} are based on matrix $L$, which leads to an extremely high computational complexity and a very complex form for controllers. Moreover, in \cite{LuJ2016IToACp1658}, the controllers are free control sequences which can take arbitrary values, but not state feedback controllers. In this paper, a new approach to design state feedback pinning control is proposed based on network structure of BNs without using the traditional state transition matrix $L$. An efficient and low dimensional control strategy is proposed based on local neighbors of controlled nodes. Motivated by the above discussions, this paper makes the following contributions: 
\begin{enumerate}
	\item[(I)] A new method named network-structure-based (NS-based) distributed pinning control is firstly proposed to achieve global stabilization. The design of NS-based distributed pinning control is based on the network structure describing the coupling connections among nodes, and the neighbors' local information. Compared with existing results on traditional $L$-based pinning control, the information of matrix $L$ is no longer needed;
	\item[(II)] The computational complexity will be dramatically reduced from $O(2^n\times 2^n)$ to $O(2\times 2^K)$, where $K$ is the largest number of in-neighbors of a node. 
	The method of designing NS-based distributed pinning control can be applied for some large-scale BNs, and this is well illustrated by several biological networks with different network sizes like 90 nodes;
	\item[(III)] By searching feedback arc sets in the network structure of BNs, pinning controlled nodes are easily determined to achieve global stability. Then, an NS-based distributed pinning control is designed based on the neighbors' local information but not on global state information. 
	This NS-based distributed pinning leads to a lower dimensional controller design and be easier for implementation than traditional methods for designing controllers;
\end{enumerate}
The summarized comparisons between NS-based distributed pinning control and traditional $L$-based pinning control can be observed from Fig. \ref{comparisons}, which also implies one challenging topic of combining network structure with state transition digraph for designing efficient pinning control.

The remainder of this paper is presented as follows: In Section \ref{preliminries}, network structure and global stability of BNs are introduced. In Section \ref{a-new-frame-section}, the main results on NS-based distributed pinning control design are presented. 
In Section \ref{simulations}, two biological networks are given to demonstrate the validity of the results.  
%Section \ref{discussions} illustrates some comparisons and the main contributions of this paper. 
A brief conclusion is given in Section \ref{conclusion}.

Some basic notations are given as follows: $\textbf{1}_{n}=(1, 1, \dots, 1)^\top$; $\mathcal {D}=\{0, 1\}$: the logical domain; $\mathbb{N}^{+}$: the set of positive integers; $\mathscr{R}_{m\times n}$: the set of $m\times n$ real matrices; $|S|$ is the cardinal number of a given set $S$; $[a, b]=\{a, a+1, \dots, b\}$, where $a<b$ and $a, b\in\mathbb{N}^{+}$; $\delta_{n}^{i}$: the $i$-th column of the identity matrix $I_{n}$; $\Delta_{n}=\{\delta_{n}^{1}, \dots, \delta_{n}^{n}\}$ denotes the columns set of $I_{n}$; $\textrm{Col}_{i}(A)$: the $i$-th column of matrix $A$. $\textrm{Col}(A)$ denotes the set of columns of $A$; $L=[\delta_{n}^{i_1}, \dots, \delta_{n}^{i_s}]$ is called a logical matrix, and we simply denote it by $L=\delta_{n}[i_1, \dots, i_s]$; $\mathcal{L}_{m\times n}$: the set of logical matrices with dimension $m\times n$; $\lfloor a \rfloor_{S}$ be the set of positive integers, $\lfloor a \rfloor_{S}=\{j: 0<j<a, j\in S\}$, where $S$ is a set of positive integers; Given variables $x_1, \dots, x_n$ and an index set $I=\{i_1, \dots, i_m\}\subseteq[1, n]$, define $[x_i]_{i\in I}=\{x_{i_1}, \dots, x_{i_m}\}$; Swap matrix $W_{[m, n]}=[I_{n}\otimes \delta_m^1, I_{n}\otimes \delta_m^2, \dots, I_{n}\otimes \delta_m^m]$; Power-reducing matrix $\Phi_{2^n}=\delta_{2^{2n}}[1, 2^{n}+2, (3-1)2^n+3, (4-1)2^n+4, \dots, (2^{n}-2)2^{n}+2^{n}-1, 2^{2{n}}]$.

\section{Network structure and Global stability}\label{preliminries}
STP of matrices is a generalization of the conventional matrix product, which deals with the case when the dimension-matching condition of matrix product is not satisfied \cite{ChengD2010pX}.

\begin{definition}[STP]\cite{ChengD2010pX}\label{definition-stp}
	Given two matrices $A\in\mathscr{R}_{n\times m}$ and $B\in\mathscr{R}_{p\times q}$, the STP of $A$ and $B$, denoted by $A\ltimes B$, is defined as: 
	$A\ltimes B=(A\otimes I_{l/m})(B\otimes I_{l/p})$,
	%	\begin{equation}\label{stp}
	%	A\ltimes B=(A\otimes I_{l/m})(B\otimes I_{l/p}),
	%	\end{equation}
	where $l$ is the least common multiple of $m$ and $p$, $\otimes$ is the Kronecker product of matrices.
\end{definition}

%\begin{definition}\cite{ChengD2010pX}\label{swap}
%	Define the swap matrix $W_{[m, n]}\in\mathcal {L}_{mn\times mn}$ as follows: $W_{[m, n]}=[I_{n}\otimes \delta_m^1, I_{n}\otimes \delta_m^2, \dots, I_{n}\otimes \delta_m^m]$. 
%	If $\sigma_1\in \Delta_{m}$ and $\sigma_2\in \Delta_{n}$, then $W_{[m,n]}\sigma_1\ltimes\sigma_2=\sigma_2\ltimes\sigma_1$. 
%	%		If $m=n$, denote $W_{[m, n]}$ by $W_{[n]}$ for convenience.
%\end{definition}
%
%\begin{definition} \cite{ChengD2010pX}\label{STP-lemma}
%	Matrix $\Phi_{2^n}$ is called power-reducing matrix for $2^n$-valued logical vectors, such that $a\ltimes a=\Phi_{2^n}\ltimes a$ for any $a\in\Delta_{2^n}$, and matrix $\Phi_{2^n}$ is described as follows: 
%	\begin{equation*}\label{phi}
%	\begin{array}{lc}
%	\Phi_{2^n}=&\delta_{2^{2n}}[1, 2^{n}+2, (3-1)2^n+3, (4-1)2^n+4, \dots, \\
%	&\dots, (2^{n}-2)\cdot2^{n}+2^{n}-1, 2^{2{n}}].
%	\end{array}
%	\end{equation*}
%\end{definition}

In order to facilitate the use of the STP method in BNs, identify ``1'' and ``0'' with vectors, $1\sim\delta_2^1$, $0\sim\delta_2^2$, respectively. Under this framework, the equivalent algebraic form for any logical function can be obtained.

\begin{lemma}\cite{ChengD2010pX}\label{prop-logical}
	Let $f(a_1, \dots, a_{n}): (\mathcal {D})^{n}\rightarrow \mathcal {D}$ be a logical function. Then for every $(a_1, \dots, a_{n})\in (\Delta_2)^{n}$, there exists a unique matrix $F\in\mathcal {L}_{2\times 2^{n}}$ such that $f(a_1, \dots, a_{n})=F\ltimes a_1\ltimes \dots \ltimes a_{n}$.
	%	\begin{equation}\label{alg-form}
	%	f(a_1, \dots, a_{n})=F\ltimes a_1\ltimes \dots \ltimes a_{n}.
	%	\end{equation}
	$F$ is called the structure matrix of $f$.
\end{lemma}

A Boolean function $f: \mathcal{D}^{n}\rightarrow\mathcal{D}$ consisting of $n$ Boolean variables $x_1, \dots, x_n\in\mathcal{D}$ is a discrete-time finite state dynamical system. Then, a BN can be described by a set of Boolean functions $f_1, \dots, f_n$: $x_i(t+1)=f_i(x_1(t), \dots, x_n(t)), i\in[1, n]$. Let $x(t)=(x_1(t), \dots, x_n(t))\in\mathcal{D}^n$ be the state at time $t$.

Note that using Boolean algebra, some variables may be nonfunctional in a logical function. For example, if $f(x_1, x_2)=(x_1\wedge x_2)\vee (x_1\wedge \neg x_2)$, then $f$ is not dependent on variable $x_2$, because $(x_1\wedge x_2)\vee (x_1\wedge \neg x_2)=x_1$; while if $f(x_1, x_2)=x_1\vee x_2$, then function $f$ is dependent on variables $x_1$ and $x_2$. Then, we introduce the dependency of logical functions on variables.

\begin{definition}[Functional Variables]\label{dependency}
	A logical function $f(x_1, \dots, x_i, \dots, x_n): \mathcal{D}^{n}\rightarrow\mathcal{D}$ is said to be dependent on variable $x_i$ if there exists a tuple $\bar{x}\in\mathcal{D}^{n-1}$ such that $f(\bar{x}, x_i)\neq f(\bar{x}, \neg x_i)$, where $\bar{x}=(x_1, \dots, x_{i-1}, x_{i+1}, \dots, x_n)$. Variable $x_i$ is called a functional variable of $f(\dots, x_i, \dots)$, otherwise, it is called a nonfunctional variable. In addition, if there is no nonfunctional variables, then $f$ is called minimally represented.
\end{definition}

Based on logical functions $f_1, \dots, f_n$, introduce the interaction digraph describing the coupling connections among nodes of a BN which is called network structure.
\begin{definition}[Interaction Digraph of BNs]\label{dependency-graph}
	Consider a BN with logical functions $f=(f_1, \dots, f_n)$. The interaction digraph is a digraph denoted by $\textrm{G}_{id}=(\textrm{V}, \textrm{E})$ of $n$ vertices, the vertices set is $\textrm{V}=\{1, \dots, n\}$. An edge $i\rightarrow j$ (simplified by $e_{ij}\in\textrm{E}$) exists in $\textrm{G}_{id}=(\textrm{V}, \textrm{E})$ if and only if $f_j$ is dependent on $x_i$. 
\end{definition}

Since a Boolean function $f: \mathcal{D}^{n}\rightarrow\mathcal{D}$ can be minimally represented, in this paper the following minimally represented BN is considered:
\begin{equation}\label{bn}
	%\left\{
	%\begin{array}{cl}
	%x_1(t+1)&=f_1([x_j(t)]_{j\in\mathcal{N}_1}), \\
	%\vdots& \\
	%x_n(t+1)&=f_n([x_j(t)]_{j\in\mathcal{N}_n}). \\
	x_i(t+1)=f_i([x_j(t)]_{j\in\mathcal{N}_i}), \quad i\in[1, n].
	%\end{array}
	%\right.
\end{equation}
This implies that for each $i\in[1, n]$, all of the variables $x_j$, $j\in\mathcal{N}_i$ are functional variables for function $f_i$. Here, $x_i\in\mathcal{D}$, $i\in[1, n]$, is the $i$-th node of system \eqref{bn}, $\mathcal{N}_i\subseteq[1, n]$, $i\in[1, n]$, is the index sets of in-neighbors of node $x_i$ expressing the adjacency relation of nodes, $f_i: \mathcal{D}^{|\mathcal{N}_i|}\rightarrow\mathcal{D}$, $i\in[1, n]$, is Boolean function. 

Then, the definitions of fixed points and cyclic attractors of system \eqref{bn} are introduced as follows:

\begin{definition}[Fixed Points and Cyclic Attractors]\label{fixed-point}
	Given a BN \eqref{bn}, $\chi\in\mathcal{D}^n$ is called a fixed point if $L\chi=\chi$, and the sequence of states $\{\delta_{2^n}^{r_0}, \delta_{2^n}^{r_1}, \dots, \delta_{2^n}^{l-1}\}\subseteq\Delta_{2^n}$ is called a cyclic attractor with length $l$ if $L^l\delta_{2^n}^{r_0}=\delta_{2^n}^{r_0}$, the elements in $\{\delta_{2^n}^{r_0}, \delta_{2^n}^{r_1}, \dots, \delta_{2^n}^{r_{l-1}}\}$ are pairwise distinct.   
\end{definition}

According to Lemma \ref{prop-logical}, suppose that the structure matrices of $f_i$ are $A_i\in\mathcal{L}_{2\times 2^{|\mathcal{N}_i|}}$, $i\in[1, n]$, then one can obtain the following algebraic form,
\begin{equation}\label{bn-alg}
	\begin{array}{l}
		x_i(t+1)=A_i\ltimes_{j\in\mathcal{N}_i}x_j(t), \quad i\in[1, n].\\
	\end{array}
\end{equation}
Suppose that for the set of in-neighbors of node $x_i$, $\mathcal{N}_i=\{\textrm{b}_j^i~|~\textrm{b}_j^i\in[1, n], j\in[1, |\mathcal{N}_i|]\}$. For the indexes $\textrm{b}_j^i\in[1, n]$, $j\in[1, |\mathcal{N}_i|]$, we assume that $1\leq \textrm{b}_1^i< \textrm{b}_2^i<\dots<\textrm{b}_{|\mathcal{N}_i|}^i\leq n$. For simplicity, we always assume that $\ltimes_{j\in\mathcal{N}_i}x_j(t)\triangleq x_{\textrm{b}_1^i}(t)\ltimes x_{\textrm{b}_2^i}(t)\ltimes\dots\ltimes x_{\textrm{b}_{|\mathcal{N}_i|}^i}(t)$. For example, let $\mathcal{N}_i=\{3, 5, 6\}$, under the ordering assumption, then $\ltimes_{j\in\mathcal{N}_i}x_j(t)\triangleq x_3(t)x_5(t)x_6(t)$.

%\begin{remark}\label{re}
%	Suppose that for the set of in-neighbors of node $x_i$, $\mathcal{N}_i=\{\textbf{b}_j^i~|~\textbf{b}_j^i\in[1, n], j\in[1, |\mathcal{N}_i|]\}$. For the indexes $\textbf{b}_j^i\in[1, n]$, $j\in[1, |\mathcal{N}_i|]$, we assume that $1\leq \textbf{b}_1^i< \textbf{b}_2^i<\dots<\textbf{b}_{|\mathcal{N}_i|}^i\leq n$. For simplicity, we always assume that $\ltimes_{j\in\mathcal{N}_i}x_j(t)\triangleq x_{\textbf{b}_1^i}(t)\ltimes x_{\textbf{b}_2^i}(t)\ltimes\dots\ltimes x_{\textbf{b}_{|\mathcal{N}_i|}^i}(t)$. For example, let $\mathcal{N}_i=\{3, 5, 6\}$, under the ordering assumption, then $\ltimes_{j\in\mathcal{N}_i}x_j(t)\triangleq x_3(t)x_5(t)x_6(t)$. 
%\end{remark}

Further, letting $x(t)=\ltimes_{j=1}^nx_j(t)\in\Delta_{2^n}$, one can obtain the equivalent representation,
\begin{equation}\label{alg-eq}
	x(t+1)=Lx(t), 
\end{equation}
where $L\in\mathcal{L}_{2^n\times 2^n}$ is called the state transition matrix of system \eqref{alg-eq}. Hence, according to matrix $L$, one can obtain the state transition graph of BN \eqref{alg-eq} consisting of $2^n$ nodes, i.e. $\{\delta_{2^n}^1, \delta_{2^n}^2, \dots, \delta_{2^n}^{2^n}\}$, and there exists an edge from $\delta_{2^n}^i$ to $\delta_{2^n}^j$ in the state transition graph if and only if $L\delta_{2^n}^i=\delta_{2^n}^j$. The detailed calculations for the state transition matrix $L$ can be obtained by using the STP method \cite{ChengD2010pX}.

\begin{remark}\label{rem-1}
	Note that the dimension of matrix $L\in\mathcal{L}_{2^n\times 2^n}$ grows exponentially with the size of network (depending on the number of nodes). Thus, all the algorithms based on matrix $L$ have exponential complexity, which is a big noteworthy drawback \cite{ZhaoY2016IToNNaLSp1527}. 
	As one can see from the discussions and analysis below, the design of NS-based distributed pinning controllers is not based on matrix $L$ but on  $A_i\in\mathcal{L}_{2\times 2^{|\mathcal{N}_i|}}$ in \eqref{bn-alg}. Generally, $|\mathcal{N}_i|\ll n$ in most gene regulatory networks, thus, the approach proposed in this paper reduces calculations and the computational complexity from $O(2^n\times 2^n)$ to $O(2\times 2^K)$, where $K\leq n$ is the largest number of in-neighbors of nodes. 
\end{remark}

Based on the algebraic form \eqref{alg-eq}, some results about global stability of BNs are recalled below. 

\begin{definition}[Global Stability]\label{global-stability-1}\cite{ChengD2010pX}\cite{robert2012discreteSSBM}
	System \eqref{bn} is said to be globally stable if there exists a unique fixed point as the attractor with no other cyclic attractors. In prarticular, system \eqref{bn} is globally stable to $x^{\ast}\in\Delta_{2^n}$, if for any initial state $x_0\in\Delta_{2^n}$, there exists an integer $T$, such that $x(t, x_0)=x^{\ast}, t\geq T$.
\end{definition}

\begin{lemma}\label{thm-graph}\cite{ChengD2010pX}\cite{robert2012discreteSSBM}
	System \eqref{bn} is globally stable if the interaction digraph $\textrm{G}_{id}=(\textrm{V}, \textrm{E})$ is acyclic. In addition, assume that stable state is $x^{\ast}=\delta_{2^n}^{\gamma}$, $\gamma\in[1, 2^n]$, then system \eqref{bn} is globally stable to $x^{\ast}=\delta_{2^n}^\gamma$ if and only if there exists an integer $T$, such that $\textrm{Col}(L^T)=\{\delta_{2^n}^\gamma\}$.
\end{lemma}

\section{Main results of NS-based distributed pinning control}\label{a-new-frame-section}
Now, in this section, a new approach to design an NS-based distributed pinning control for global stabilization is proposed, based on the transformation of certain structure matrices of several nodes of BN \eqref{bn}, under which the interaction digraph $\textrm{G}_{id}=(\textrm{V}, \textrm{E})$ will become acyclic. Hence, two questions should be answered in order to achieve global stabilization to a given fixed point, one is how to determine the pinning controlled nodes, another one is how to transform the structure matrices of the controlled nodes. 
Then, the following two main steps are introduced to answer the above two questions for designing NS-based distributed pinning control, here we assume that the given fixed point is $\delta_{2^n}^\gamma$, $\gamma\in[1, 2^n]$: 
\begin{enumerate}
	\item[(1)] use the concept of feedback arc set to determine the controlled nodes (see Problem 1 below), and delete the minimum number of edges in the interaction digraph of BN \eqref{bn} based on the transformation of structure matrices of controlled nodes (see Problem 2 below), such that $\textrm{G}_{id}=(\textrm{V}, \textrm{E})$ becomes acyclic, which guarantees global stabilization but not to the given fixed point; 
	\item[(2)] after guaranteeing global stabilization, use an integer linear programming problem (see Problem 3 below) to further determine controlled nodes that are needed to achieve the given fixed point $\delta_{2^n}^\gamma$, and further transform the structure matrices $A_1, \dots, A_n$, so that state $\delta_{2^n}^\gamma$ is the unique fixed point. 
\end{enumerate}
It should be noted that the control action of deleting edges of interaction digraph $\textrm{G}_{id}=(\textrm{V}, \textrm{E})$ is plausible in genetic networks. In \cite{Murrugarra2016BSBp94}, two types of control actions (one is deletion of edges and another one is deletion of nodes) have been applied to Boolean molecular networks. For example, the deletion of an edge can be achieved by the use of therapeutic drugs to control some specific gene interactions. In the following subsections, we will use these two steps to design an NS-based distributed pinning control to achieve global stabilization to any given fixed point based on the interaction digraph $\textrm{G}_{id}=(\textrm{V}, \textrm{E})$ of BN \eqref{bn}.

\subsection{\textbf{Step 1: guarantee global stabilization} }  

%\footnotetext[1]{Depth-first search algorithm is a fundamental recursive algorithm following the edges of a digraph to find vertices connected to the source \cite{Sedgewick2011AWP}. }

Now, we consider the first main step, that is to delete the minimum number of edges in the interaction digraph $\textrm{G}_{id}=(\textrm{V}, \textrm{E})$, such that the reduced digraph is acyclic. 
By using the depth-first search algorithm \cite{Sedgewick2011AWP}, one can obtain the cycles and fixed points in the digraph $\textrm{G}_{id}=(\textrm{V}, \textrm{E})$. Firstly, we present some notations. Given an edge $e\in \textrm{E}$ in $\textrm{G}_{id}=(\textrm{V}, \textrm{E})$, let $\textrm{O}_{-}(e)$ be the starting vertex and $\textrm{O}_{+}(e)$ be the ending vertex of the edge $e\in E$. Then, the concept of (minimum) feedback arc set for a given digraph is introduced in order to determine the pinning controlled nodes.

\begin{definition}[Feedback Arc Set]\cite{Bang2008SSBM}\label{fas}
	Given a directed network, a feedback arc set is a subset of edges containing at least one edge of every cycle, and it is called a minimum feedback arc set if its cardinality is minimum. Therefore, the removal of a feedback arc set renders the network acyclic. 
\end{definition}

%\begin{remark}
%	Feedback arc set along with feedback vertex set are two important concepts in graph theory, which have been widely studied. In \cite{Baharev2015UVp35}, an exact method to determine the minimum feedback arc set has been proposed for large and sparse digraphs, based on linear programming method. In \cite{Baharev2015UVp35}, Baharev \emph{et al.} tested several different sizes of digraph to find a minimal feedback arc set. 
%\end{remark}

Using the depth-first search algorithm, one can obtain all of the minimal feedback arc sets in the interaction digraph $\textrm{G}_{id}=(\textrm{V}, \textrm{E})$, based on which, the following problem is considered to determine the pinning controlled nodes. 

\textbf{Problem 1 (Determination of Pinning Nodes):} Assume that $\Omega_1=\{e_1^1, \dots, e_{\textbf{c}_2}^1\}, \dots, \Omega_\kappa=\{e_1^\kappa, \dots, e_{\textbf{c}_2}^\kappa\}$ are all of the minimal feedback arc sets with cardinality $\textbf{c}_2\in[1, |\textrm{E}|]$ in the interaction digraph $\textrm{G}_{id}=(\textrm{V}, \textrm{E})$. Here, $e_1^1, \dots, e_{\textbf{c}_2}^1$, $\dots$, $e_1^\kappa, \dots, e_{\textbf{c}_2}^\kappa\in \textrm{E}$, $\kappa\in\mathbb{N}^{+}$. Then, minimize the cost function
\begin{equation}\label{card-cost-1}
	\textbf{c}_1\triangleq\textrm{min}\{|\pi_j|: j=1, \dots, \kappa\}, 
\end{equation}
subject to conditions
\begin{equation}\label{card-cost-2}
	\pi_j\triangleq\bigcup_{i=1}^{\textbf{c}_2}\textrm{O}_{+}(e_i^{j}), j\in[1, \kappa].
\end{equation}

\begin{remark}
	As one can see from Problem 1, \eqref{card-cost-1} and \eqref{card-cost-2} are presented to find the minimum number of ending vertices among the edges in the minimal feedback arc sets, such that the digraph $\textrm{G}_{id}=(\textrm{V}, \textrm{E})$ becomes acyclic if the edges in the minimal feedback arc sets are deleted. 
	%Actually, \eqref{card-cost-2} is just to find a minimum feedback arc set in the interaction digraph of BNs. 
	%A feedback acr set is a subset of edges containing at least one edge of every cycle in a digraph \cite{Bang2008SSBM}. 
	%Finding a feedback acr set of minimum cardinality is the minimum feedback arc set problem. 
	During the past few years, much attention has been focused on finding the relationship between fixed points and the minimum feedback arc set in interaction digraph of BNs \cite{Mori2017PRLp028301}. 
	%When the network is not sparse, it will be a challenging problem to find a minimum feedback arc set. 
	In general, many real-world genetic regulatory networks are sparse. Thus, finding a minimum feedback arc set satisfying \eqref{card-cost-2} is feasible in practice. After considering Problem 1, the controlled nodes can be easily determined, which are the ending vertices of edges in the minimal feedback arc sets in the interaction digraph of \eqref{bn}.
\end{remark}

Assume that $\Omega_\varsigma=\{e_1^\varsigma, \dots, e_{\textbf{c}_2}^\varsigma\}\in\{\Omega_1, \dots, \Omega_\kappa\}$ ($\varsigma\in[1, \kappa]$) is one 
feasible minimal feedback arc set with cardinality $\textbf{c}_2$, which satisfies conditions \eqref{card-cost-1} and \eqref{card-cost-2}. Then, we further assume
\begin{equation}\label{121212}
	\bigcup_{i=1}^{\textbf{c}_2}\textrm{O}_{+}(e_i^{\varsigma})\triangleq\{\omega_1, \dots, \omega_{\textbf{c}_1}\}, ~~\omega_1, \dots, \omega_{\textbf{c}_1}\in\textrm{V}, 
\end{equation}
which are the ending vertices of edges of the minimal feedback arc set $\Omega_\varsigma=\{e_1^\varsigma, \dots, e_{\textbf{c}_2}^\varsigma\}$.
Thus, under these assumptions, nodes $x_i$, $i\in\{\omega_1, \dots, \omega_{\textbf{c}_1}\}$, should be controlled after considering Problem 1. Since some edges may share a same ending vertices, \eqref{121212} implies that vertices $\omega_1, \dots, \omega_{\textbf{c}_1}$ are all the possible ending vertices for the edges $e_1^\varsigma, \dots, e_{\textbf{c}_2}^\varsigma$. Here, we assume that the edges $e_{l_1^1}^\varsigma, \dots, e_{l_1^{\varepsilon_1}}^\varsigma\in\{e_1^\varsigma, \dots, e_{\textbf{c}_2}^\varsigma\}$ ($l_1^1, \dots, l_1^{\varepsilon_1}\in[1, \textbf{c}_2]$) share the same ending vertex $\omega_1$, i.e. $\bigcup_{j=l_1^1}^{l_1^{\varepsilon_1}}\textrm{O}_{+}(e_j^\varsigma)\triangleq\{\omega_1\}$; $\dots$; the edges $e_{l_{\textbf{c}_1}^1}^\varsigma, \dots, e_{l_{\textbf{c}_1}^{\varepsilon_{\textbf{c}_1}}}^\varsigma\in\{e_1^\varsigma, \dots, e_{\textbf{c}_2}^\varsigma\}$ ($l_{\textbf{c}_1}^1, \dots, l_{\textbf{c}_1}^{\varepsilon_{\textbf{c}_1}}\in[1, \textbf{c}_2]$) share the same ending vertex $\omega_{\textbf{c}_1}$, i.e. $\bigcup_{j=l_{\textbf{c}_1}^1}^{l_{\textbf{c}_1}^{\varepsilon_{\textbf{c}_1}}}\textrm{O}_{+}(e_j^\varsigma)\triangleq\{\omega_{\textbf{c}_1}\}$. In addition, $\bigcup_{j=l_1^1}^{l_1^{\varepsilon_1}}\textrm{O}_{-}(e_j^\varsigma)\triangleq\{\theta_1^{1}, \dots, \theta_1^{\varepsilon_1}\}$, $\dots$, $\bigcup_{j=l_{\textbf{c}_1}^1}^{l_{\textbf{c}_1}^{\varepsilon_{\textbf{c}_1}}}\textrm{O}_{-}(e_j^\varsigma)\triangleq\{\theta_{\textbf{c}_1}^{1}, \dots, \theta_{\textbf{c}_1}^{\varepsilon_{\textbf{c}_1}}\}$. Here, assume that $\theta_1^{1}, \dots, \theta_1^{\varepsilon_1}$, $\dots$, $\theta_{\textbf{c}_1}^{1}$, $\dots$, $\theta_{\textbf{c}_1}^{\varepsilon_{\textbf{c}_1}}\in\textrm{V}$, and $\theta_1^{1}< \dots< \theta_1^{\varepsilon_1}$, $\dots$, $\theta_{\textbf{c}_1}^{1}<\dots<\theta_{\textbf{c}_1}^{\varepsilon_{\textbf{c}_1}}$.

After determining the controlled nodes based on the minimal feedback arc set given in Problem 1, the following problem is considered to further transform corresponding structure matrices of controlled nodes $x_i$, $i\in\{\omega_1, \dots, \omega_{\textbf{c}_1}\}$ for designing NS-based distributed pinning control. 
%Note that structure matrix is one-to-one mapping to the corresponding logical function, thus the following problem is considered to transform certain logical functions on the above found .

\textbf{Problem 2 (Transformation of Structure Matrices of Pinning Nodes):} Consider the nodes $x_i$, $i\in\{\omega_1, \dots, \omega_{\textbf{c}_1}\}$ that should be controlled, find matrices $\tilde{A}_{\omega_1}\in\mathcal{L}_{2\times 2^{|\mathcal{N}_{\omega_1}|}}$, $\dots$, $\tilde{A}_{\omega_{\textbf{c}_1}}\in\mathcal{L}_{2\times 2^{|\mathcal{N}_{\omega_{{\textbf{c}_1}}}|}}$, and $\hat{A}_{\omega_1}\in\mathcal{L}_{2\times 2^{|\bar{\mathcal{N}}_{\omega_1}|}}, \dots, \hat{A}_{\omega_{\textbf{c}_1}}\in\mathcal{L}_{2\times 2^{|\bar{\mathcal{N}}_{\omega_{{\textbf{c}_1}}}|}}$, such that 
\begin{equation}\label{pro-eq-constraint}
	\left\{
	\begin{array}{rl}
		\tilde{A}_{\omega_1}&=\hat{A}_{\omega_1}(I_{2^{|\bar{\mathcal{N}}_{\omega_1}|}}\otimes \textbf{1}_{2^{|\mathcal{N}_{\omega_1}^{c}|}}^\top), \\
		&\cdots \\
		\tilde{A}_{\omega_{\textbf{c}_1}}&=\hat{A}_{\omega_{\textbf{c}_1}}(I_{2^{|\bar{\mathcal{N}}_{\omega_{\textbf{c}_1}}|}}\otimes \textbf{1}_{2^{|\mathcal{N}_{\omega_{\textbf{c}_1}}^{c}|}}^\top).
	\end{array}
	\right.
\end{equation}
Here, we denote $\bar{\mathcal{N}}_{\omega_1}\triangleq\mathcal{N}_{\omega_1}\backslash\{\theta_1^{1}, \dots, \theta_1^{\varepsilon_1}\}$, $\dots$, $\bar{\mathcal{N}}_{\omega_{\textbf{c}_1}}\triangleq\mathcal{N}_{\omega_{\textbf{c}_1}}\backslash\{\theta_{\textbf{c}_1}^{1}, \dots, \theta_{\textbf{c}_1}^{\varepsilon_{\textbf{c}_1}}\}$, and $\mathcal{N}_{\omega_1}^{c}\triangleq\{\theta_1^{1}, \dots, \theta_1^{\varepsilon_1}\}$, $\dots$, $\mathcal{N}_{\omega_{\textbf{c}_1}}^{c}\triangleq\{\theta_{\textbf{c}_1}^{1}, \dots, \theta_{\textbf{c}_1}^{\varepsilon_{\textbf{c}_1}}\}$. 

For example, consider matrix $\tilde{A}_{\omega_1}$ satisfying \eqref{pro-eq-constraint}, i.e. $\tilde{A}_{\omega_1}=\hat{A}_{\omega_1}(I_{2^{|\bar{\mathcal{N}}_{\omega_1}|}}\otimes \textbf{1}_{2^{|\mathcal{N}_{\omega_1}^{c}|}}^\top)$, then using STP method, suppose that the algebraic form of node $x_{\omega_1}$ becomes $x_{\omega_1}(t+1)=\tilde{A}_{\omega_1}\ltimes_{j\in\bar{\mathcal{N}}_{\omega_j}}x_j(t)\ltimes_{j\in\mathcal{N}^{c}_{\omega_j}}x_j(t)=\hat{A}_{\omega_1}(I_{2^{|\bar{\mathcal{N}}_{\omega_1}|}}\otimes \textbf{1}_{2^{|\mathcal{N}_{\omega_1}^{c}|}}^\top)\ltimes_{j\in\bar{\mathcal{N}}_{\omega_j}}x_j(t)\ltimes_{j\in\mathcal{N}^{c}_{\omega_j}}x_j(t)$, one can finally simplify as follows: $x_{\omega_1}(t+1)=\hat{A}_{\omega_1}\ltimes_{j\in\bar{\mathcal{N}}_{\omega_1}}x_j(t)$. This implies that in the interaction digraph $\textrm{G}_{id}=(\textrm{V}, \textrm{E})$, the corresponding edges from nodes $x_{\theta_1^{1}}, \dots, x_{\theta_1^{\varepsilon_1}}$ to node $x_{\omega_1}$ will be deleted. Hence, consider the algebraic forms of nodes $x_{\omega_1}, \dots, x_{\omega_{\textbf{c}_1}}$: $x_{\omega_1}(t+1)=A_{\omega_1}\ltimes_{j\in\mathcal{N}_{\omega_1}}x_j(t), \dots, x_{\omega_{\textbf{c}_1}}(t+1)=A_{\omega_{\textbf{c}_1}}\ltimes_{j\in\mathcal{N}_{\omega_{\textbf{c}_1}}}x_j(t)$, after finding matrices $\tilde{A}_{\omega_1}$, $\dots$, $\tilde{A}_{\omega_{\textbf{c}_1}}$, and $\hat{A}_{\omega_1}, \dots, \hat{A}_{\omega_{\textbf{c}_1}}$ satisfying \eqref{pro-eq-constraint}, if the structure matrices $A_{\omega_1}, \dots, A_{\omega_{\textbf{c}_1}}$ are transformed to matrices $\tilde{A}_{\omega_1}, \dots, \tilde{A}_{\omega_{\textbf{c}_1}}$, then the edges in the minimal feedback arc sets found in Problem 1 will be deleted in the digraph $\textrm{G}_{id}=(\textrm{V}, \textrm{E})$ which leads to an acyclic network structure. Note that there are many feasible matrices $\tilde{A}_{\omega_1}$, $\dots$, $\tilde{A}_{\omega_{\textbf{c}_1}}$, and $\hat{A}_{\omega_1}, \dots, \hat{A}_{\omega_{\textbf{c}_1}}$ satisfying \eqref{pro-eq-constraint}, one can firstly choose matrices $\hat{A}_{\omega_1}, \dots, \hat{A}_{\omega_{\textbf{c}_1}}$ in order to determine $\tilde{A}_{\omega_1}$, $\dots$, $\tilde{A}_{\omega_{\textbf{c}_1}}$ satisfying \eqref{pro-eq-constraint}. Since matrices $\hat{A}_{\omega_1}, \dots, \hat{A}_{\omega_{\textbf{c}_1}}$ are logical matrices with finite dimensions, there only exist finite possible cases for $\hat{A}_{\omega_1}, \dots, \hat{A}_{\omega_{\textbf{c}_1}}$ .

Using the property of swap matrix $W_{[m, n]}$, that is $W_{[m,n]}\sigma_1\ltimes\sigma_2=\sigma_2\ltimes\sigma_1$, for $x_{\omega_j}(t+1)=A_{\omega_j}\ltimes_{j\in\mathcal{N}_{\omega_j}}x_j(t)$, $j\in[1, \textbf{c}_1]$, one can swap the positions of neighbors as follows:
\begin{equation}\label{change}
	\begin{array}{rl}
		x_{\omega_j}(t+1)=&A_{\omega_j}\textbf{W}_j\ltimes_{j\in\bar{\mathcal{N}}_{\omega_j}}x_j(t)\ltimes_{j\in\mathcal{N}^{c}_{\omega_j}}x_j(t), \\
		\triangleq& \bar{A}_{\omega_j}\ltimes_{j\in\bar{\mathcal{N}}_{\omega_j}}x_j(t)\ltimes_{j\in\mathcal{N}^{c}_{\omega_j}}x_j(t), \\
	\end{array}
\end{equation}
where $\textbf{W}_j\triangleq[\ltimes_{i=1}^{\omega_j}W_{[2, 2^{|\lfloor \theta_j^{i}\rfloor_{\mathcal{N}_{\omega_j}}|}]}W_{[2^{|\mathcal{N}^c_{\omega_j}|}, 2^{|\mathcal{N}_{\omega_j}|}]}]\otimes I_{2^{|\mathcal{N}_{\omega_j}|-|\mathcal{N}^c_{\omega_j}|}}$ are invertible matrices, and $\bar{A}_{\omega_j}\triangleq A_{\omega_j}\textbf{W}_j$.

Then, based on Problems 1 and 2, one can design an NS-based distributed pinning control to guarantee global stabilization on nodes $x_j$, $j\in\{\omega_1, \dots, \omega_{\textbf{c}_1}\}$, in the following form:
\begin{equation}\label{pinning-controllers-step-1}
	\left\{
	\begin{array}{rl}
		x_j(t+1)=&u_j(t)\oplus_jf_j([x_{j_1}(t)]_{j_1\in\mathcal{N}_j}), j\in\{\omega_1, \dots, \omega_{\textbf{c}_1}\}, \\
		u_j(t)=&g_j([x_{j_1}(t)]_{j_1\in\mathcal{N}_j}), ~~~~~~~~~~j\in\{\omega_1, \dots, \omega_{\textbf{c}_1}\}, \\
		x_i(t+1)=&f_i([x_{j_2}(t)]_{j_2\in\mathcal{N}_i}), ~~~~i\in[1, n]\backslash\{\omega_1, \dots, \omega_{\textbf{c}_1}\}. \\
	\end{array}
	\right.
\end{equation}
Here, $u_j\in\mathcal{D}$, $j\in\{\omega_1, \dots, \omega_{\textbf{c}_1}\}$, are NS-based distributed pinning controllers for nodes $x_j$, $j\in\{\omega_1, \dots, \omega_{\textbf{c}_1}\}$. $\oplus_j: \mathcal{D}^2\rightarrow\mathcal{D}$, $j\in\{\omega_1, \dots, \omega_{\textbf{c}_1}\}$, are binary logical functions to be determined in the following sequel. Functions $g_j :\mathcal{D}^{|\mathcal{N}_j|}\rightarrow\mathcal{D}$, $j\in\{\omega_1, \dots, \omega_{\textbf{c}_1}\}$, are Boolean functions of controllers $u_j$ depending on the corresponding neighbors of nodes $x_j$ ($j\in\{\omega_1, \dots, \omega_{\textbf{c}_1}\}$), which will be determined later. Since any logical function can be equivalently expressed by means of suitable structure matrix, let $M_{\oplus_j}\in\mathcal{L}_{2\times 4}$ and $K_j\in\mathcal{L}_{2\times 2^{|\mathcal{N}_{j}|}}$, $j\in\{\omega_1, \dots, \omega_{\textbf{c}_1}\}$, denote the structure matrices of functions $\oplus_j: \mathcal{D}^2\rightarrow\mathcal{D}$ and $g_j :\mathcal{D}^{|\mathcal{N}_j|}\rightarrow\mathcal{D}$. Thus, in the following, one needs to determine the structure matrices $M_{\oplus_j}$ and $K_j$, to determine the logical functions $\oplus_j: \mathcal{D}^2\rightarrow\mathcal{D}$ and $g_j :\mathcal{D}^{|\mathcal{N}_j|}\rightarrow\mathcal{D}$.

Let $\textbf{K}_{j}\triangleq K_{j}\textbf{W}_j\in\mathcal{L}_{2\times 2^{|\mathcal{N}_{j}|}}$, $j\in\{\omega_1, \dots, \omega_{\textbf{c}_1}\}$. For nodes $x_j$, $j\in\{\omega_1, \dots, \omega_{\textbf{c}_1}\}$, using STP, Eq. \eqref{pinning-controllers-step-1} can be expressed as follows:
\begin{equation}\label{step-controlled}
	\left\{
	\begin{array}{l}
		x_j(t+1)=M_{\oplus_j}u_j(t)\bar{A}_{j}\ltimes_{i\in\bar{\mathcal{N}}_{j}}x_i(t)\ltimes_{i\in\mathcal{N}^{c}_{j}}x_i(t), \\
		~~=M_{\oplus_j}\textbf{K}_j(I_{2^{|\mathcal{N}_j|}}\otimes \bar{A}_j)\Phi_{2^{|\mathcal{N}_j|}}\ltimes_{i\in\bar{\mathcal{N}}_{j}}x_i(t)\ltimes_{i\in\mathcal{N}^{c}_{j}}x_i(t), \\
		u_j(t)=\textbf{K}_{j}\ltimes_{i\in\bar{\mathcal{N}}_{j}}x_i(t)\ltimes_{i\in\mathcal{N}^{c}_{j}}x_i(t).
	\end{array}
	\right.
\end{equation}
According to Problems 1 and 2, suppose that for node $x_j$, $j\in\{\omega_1, \dots, \omega_{\textbf{c}_1}\}$, system $x_j(t+1)=M_{\oplus_j}\textbf{K}_j(I_{2^{|\mathcal{N}_j|}}\otimes \bar{A}_j)\Phi_{2^{|\mathcal{N}_j|}}\ltimes_{i\in\bar{\mathcal{N}}_{j}}x_i(t)\ltimes_{i\in\mathcal{N}^{c}_{j}}x_i(t)$ becomes $x_j(t+1)=\hat{A}_{j}(I_{2^{|\bar{\mathcal{N}}_{j}|}}\otimes \textbf{1}_{2^{|\mathcal{N}_{j}^{c}|}}^\top)\ltimes_{i\in\bar{\mathcal{N}}_{j}}x_i(t)\ltimes_{i\in\mathcal{N}^{c}_{j}}x_i(t)$, then one can obtain $x_{j}(t+1)=\hat{A}_{j}\ltimes_{i\in\bar{\mathcal{N}}_{j}}x_i(t)$. Thus, according to $x_{j}(t+1)=\hat{A}_{j}\ltimes_{i\in\bar{\mathcal{N}}_{j}}x_i(t)$ ($j\in\{\omega_1, \dots, \omega_{\textbf{c}_1}\}$), in the interaction digraph $\textrm{G}_{id}=(\textrm{V}, \textrm{E})$, the edges from nodes $x_{\nu_1^{1}}, \cdots, x_{\nu_1^{\varepsilon_1}}$ to node $x_{\omega_1}$, $\cdots$, and the edges from nodes $x_{\nu_{\textbf{c}_1}^{1}}, \cdots, x_{\nu_{\textbf{c}_1}^{\varepsilon_{\textbf{c}_1}}}$ to node $x_{\omega_{\textbf{c}_1}}$ will be deleted. This implies that if there exist matrices $M_{\oplus_j}\in\mathcal{L}_{2\times 4}$ and $K_j\in\mathcal{L}_{2\times 2^{|\mathcal{N}_{j}|}}$, $j\in\{\omega_1, \dots, \omega_{\textbf{c}_1}\}$ satisfying $\tilde{A}_{\omega_j}=M_{\oplus_{\omega_j}}\textbf{K}_{\omega_j}(I_{2^{|\mathcal{N}_{\omega_j}|}}\otimes \bar{A}_{\omega_j})\Phi_{2^{|\mathcal{N}_{\omega_j}|}}$, then the interaction digraph $\textrm{G}_{id}=(\textrm{V}, \textrm{E})$ will become acyclic.

In order to design NS-based distributed pinning controllers, one needs to solve matrices $M_{\oplus_j}\in\mathcal{L}_{2\times 4}$ and $K_j\in\mathcal{L}_{2\times 2^{|\mathcal{N}_{j}|}}$, $j\in\{\omega_1, \dots, \omega_{\textbf{c}_1}\}$ satisfying the equations below,
\begin{equation}\label{pro-eq-constraint-22}
	\left\{
	\begin{array}{rl}
		\tilde{A}_{\omega_1}&=M_{\oplus_{\omega_1}}\textbf{K}_{\omega_1}(I_{2^{|\mathcal{N}_{\omega_1}|}}\otimes \bar{A}_{\omega_1})\Phi_{2^{|\mathcal{N}_{\omega_1}|}}, \\
		&\cdots \\
		\tilde{A}_{\omega_{\textbf{c}_1}}&=M_{\oplus_{\omega_{\textbf{c}_1}}}\textbf{K}_{\omega_{\textbf{c}_1}}(I_{2^{|\mathcal{N}_{\omega_{\textbf{c}_1}}|}}\otimes \bar{A}_{\omega_{\textbf{c}_1}})\Phi_{2^{|\mathcal{N}_{\omega_{\textbf{c}_1}}|}}.
	\end{array}
	\right.
\end{equation}
Thus, the solvability of \eqref{pro-eq-constraint-22} is vital for designing NS-based distributed pinning controllers, which is guaranteed by the following theorem.

\begin{theorem}\label{solvability}
	Given matrices $\tilde{A}_{\omega_j}$ and $\bar{A}_{\omega_j}$, $j\in[1, \textbf{c}_1]$, one can obtain that Eq. \eqref{pro-eq-constraint-22} is always solvable. 
\end{theorem}

\textbf{Proof.} Here, for simplicity, we only prove the solvability of equation $\tilde{A}_{\omega_1}=M_{\oplus_{\omega_1}}\textbf{K}_{\omega_1}(I_{2^{|\mathcal{N}_{\omega_1}|}}\otimes \bar{A}_{\omega_1})\Phi_{2^{|\mathcal{N}_{\omega_1}|}}$, the rest of \eqref{pro-eq-constraint-22} can be similarly derived. Note that 
%\begin{equation*}
%\begin{array}{rl}
$\tilde{A}_{\omega_1}=M_{\oplus_{\omega_1}}\textbf{K}_{\omega_1}(I_{2^{|\mathcal{N}_{\omega_1}|}}\otimes \bar{A}_{\omega_1})\Phi_{2^{|\mathcal{N}_{\omega_1}|}}=M_{\oplus_{\omega_1}}(\textbf{K}_{\omega_1}\otimes \bar{A}_{\omega_1})\Phi_{2^{|\mathcal{N}_{\omega_1}|}}$. 
%\end{array}
%\end{equation*}
For simplicity, here, denote $|\mathcal{N}_{\omega_1}|=n_1$, and assume that matrix
$
M_{\oplus_{\omega_1}}=\left[\begin{array}{cccc}
\alpha_1 &\alpha_2 &\alpha_3&\alpha_4 \\
1-\alpha_1 &1-\alpha_2 &1-\alpha_3&1-\alpha_4 \\
\end{array}\right]\in\mathcal{L}_{2\times 4}
$, matrix
$
\textbf{K}_{\omega_1}=\left[\begin{array}{cccc}
\beta_1 &\beta_2 &\cdots&\beta_{2^{n_1}} \\
1-\beta_1 &1-\beta_2 &\cdots&1-\beta_{2^{n_1}} \\
\end{array}\right]\in\mathcal{L}_{2\times 2^{n_1}}
$, matrix
$
\bar{A}_{\omega_1}=\left[\begin{array}{cccc}
y_1 &y_2 &\cdots&y_{2^{n_1}} \\
1-y_1 &1-y_2 &\cdots&1-y_{2^{n_1}} \\
\end{array}\right]\in\mathcal{L}_{2\times 2^{n_1}}
$, and matrix
$
\tilde{A}_{\omega_1}=\left[\begin{array}{cccc}
a_1 &a_2 &\cdots&a_{2^{n_1}} \\
1-a_1 &1-a_2 &\cdots&1-a_{2^{n_1}} \\
\end{array}\right]\in\mathcal{L}_{2\times 2^{n_1}},
$
where $\alpha_1, \alpha_2, \alpha_3, \alpha_4\in\mathcal {D}$, $\beta_j, y_j, a_j\in\mathcal {D}$, $j\in[1, 2^{n_1}]$.
Then, one can obatin the following equation,
\begin{equation}\label{eq-solvability}
	{\begin{array}{l}
			\tilde{A}_{\omega_1}=M_{\oplus_{\omega_1}}\textbf{K}_{\omega_1}(I_{2^{|\mathcal{N}_{\omega_1}|}}\otimes \bar{A}_{\omega_1})\Phi_{2^{|\mathcal{N}_{\omega_1}|}} \\
			=\left[\begin{array}{cccc}
				\alpha_1 &\alpha_2 &\alpha_3&\alpha_4 \\
				1-\alpha_1 &1-\alpha_2 &1-\alpha_3&1-\alpha_4 \\
			\end{array}\right]\times\\
			~\left[\begin{array}{cccc}
				\beta_1y_1 &\cdots&\beta_{2^{n_1}}y_{2^{n_1}} \\
				\beta_1(1-y_1) &\cdots&\beta_{2^{n_1}}(1-y_{2^{n_1}}) \\
				(1-\beta_1)y_1 &\cdots&(1-\beta_{2^{n_1}})y_{2^{n_1}} \\
				(1-\beta_1)(1-y_1) &\cdots&(1-\beta_{2^{n_1}})(1-y_{2^{n_1}}) \\
			\end{array}\right]\Phi_{2^{n_1}}\\
			=\left[\begin{array}{cccc}
				a_1 &a_2 &\cdots&a_{2^{n_1}} \\
				1-a_1 &1-a_2 &\cdots&1-a_{2^{n_1}} \\
			\end{array}\right].
		\end{array}}
	\end{equation}
	By simplifying \eqref{eq-solvability}, the solvability of equation $\tilde{A}_{\omega_1}=M_{\oplus_{\omega_1}}\textbf{K}_{\omega_1}(I_{2^{|\mathcal{N}_{\omega_1}|}}\otimes \bar{A}_{\omega_1})\Phi_{2^{|\mathcal{N}_{\omega_1}|}}$ is converted to the following equations:
	\begin{equation}\label{solve-eq}
		%\left\{
		\begin{array}{l}
			%\alpha_1\beta_1y_1+\alpha_2\beta_1(1-y_1)+\alpha_3(1-\beta_1)y_1+\\
			%\qquad\qquad\qquad\qquad\qquad\qquad\alpha_4(1-\beta_1)(1-y_1)=a_1, \\
			\alpha_1\beta_iy_i+\alpha_2\beta_i(1-y_i)+\alpha_3(1-\beta_i)y_i+\\
			\qquad\qquad\qquad~\alpha_4(1-\beta_i)(1-y_i)=a_i, ~~i\in[1, 2^{n_1}].\\
			%\cdots \\
			%\alpha_1\beta_{2^{n_1}}y_{2^{n_1}}+\alpha_2\beta_{2^{n_1}}(1-y_{2^{n_1}})+\alpha_3(1-\beta_{2^{n_1}})y_{2^{n_1}}+\\
			%\qquad\qquad\qquad\qquad\qquad\alpha_4(1-\beta_{2^{n_1}})(1-y_{2^{n_1}})=a_{2^{n_1}}. \\
		\end{array}
		%\right.
	\end{equation}
	Since matrices $\tilde{A}_{\omega_1}$ and $\bar{A}_{\omega_1}$ are known, we only need to consider the following four cases to determine the values of $\alpha_i$ and $\beta_i$ according to different values of $y_i$ and $a_i$. 
	
	Consider \eqref{solve-eq}, as for case 1: $a_i=1, y_i=1$, then \eqref{solve-eq} becomes $\alpha_1\beta_i+\alpha_3(1-\beta_i)=1$. Then, $\alpha_1\beta_i+\alpha_3(1-\beta_i)=1$ holds under solutions $\beta_i=1$ and $\alpha_1=1$ for any $\alpha_2, \alpha_3, \alpha_4\in\{1, 0\}$. 
	
	As for case 2: $a_i=1, y_i=0$, \eqref{solve-eq} holds under solution $\beta_i=0, \alpha_4=1$ for any $\alpha_2, \alpha_3, \alpha_1\in\{1, 0\}$.
	
	As for case 3: $a_i=0, y_i=1$,  \eqref{solve-eq} holds under solution $\beta_i=0, \alpha_3=0$ for any $\alpha_1, \alpha_2, \alpha_4\in\{1, 0\}$.
	
	As for case 4: $a_i=0, y_i=0$,  \eqref{solve-eq} holds under solution $\beta_i=1, \alpha_2=0$ for any $\alpha_1, \alpha_3, \alpha_4\in\{1, 0\}$.
	Thus, according to the above analysis, equation $\tilde{A}_{\omega_1}=M_{\oplus_{\omega_1}}\textbf{K}_{\omega_1}(I_{2^{|\mathcal{N}_{\omega_1}|}}\otimes \bar{A}_{\omega_1})\Phi_{2^{|\mathcal{N}_{\omega_1}|}}$ is always solvable. Then, similarly, one can also conclude that Eq. \eqref{pro-eq-constraint-22} is solvable, which completes the proof. ~\IEEEQED

Based on the solvability of \eqref{pro-eq-constraint-22}, global stabilization of BN under an NS-based distributed pinning control in the form of Eq. \eqref{pinning-controllers-step-1} will be guaranteed by the following theorem.
	
	\begin{theorem}\label{thm-global-stable}
		By solving matrices $M_{\oplus_j}\in\mathcal{L}_{2\times 4}$ and $K_j\in\mathcal{L}_{2\times 2^{|\mathcal{N}_{j}|}}$, $j\in\{\omega_1, \dots, \omega_{\textbf{c}_1}\}$ from Eq. \eqref{pro-eq-constraint-22}, BN under an NS-based distributed pinning control in the form of Eq. \eqref{pinning-controllers-step-1} will be globally stabilized. 
	\end{theorem}
	%\begin{proof}
	\textbf{Proof.} If BN \eqref{bn} is under the NS-based distributed pinning control in the form of \eqref{pinning-controllers-step-1}, then one can obtain its algebraic form for the controlled nodes $x_j$, $j\in\{\omega_1, \dots, \omega_{\textbf{c}_1}\}$, 
	\begin{equation*}\label{proof}\hspace{-2mm}
		%	\left\{
		\begin{array}{ll}
			x_j(t+1)&=M_{\oplus_j}u_j(t)\bar{A}_{j}\ltimes_{i\in\bar{\mathcal{N}}_{j}}x_i(t)\ltimes_{i\in\mathcal{N}^{c}_{j}}x_i(t), \\
			&=M_{\oplus_j}\textbf{K}_j(I_{2^{|\mathcal{N}_j|}}\otimes \bar{A}_j)\Phi_{2^{|\mathcal{N}_j|}}\ltimes_{i\in\bar{\mathcal{N}}_{j}}x_i(t)\ltimes_{i\in\mathcal{N}^{c}_{j}}x_i(t).\\
			%	u_j(t)=\textbf{K}_{j}\ltimes_{i\in\bar{\mathcal{N}}_{j}}x_i(t)\ltimes_{i\in\mathcal{N}^{C}_{j}}x_i(t).
		\end{array}
		%	\right.
	\end{equation*}
	Since matrices $M_{\oplus_j}\in\mathcal{L}_{2\times 4}$ and $K_j\in\mathcal{L}_{2\times 2^{|\mathcal{N}_{j}|}}$, $j\in\{\omega_1, \dots, \omega_{\textbf{c}_1}\}$ can be obtained by solving Eq. \eqref{pro-eq-constraint-22}, one has
	\begin{equation*}\label{proof-2}
		%	\left\{
		%\begin{array}{l}
		x_j(t+1)=\tilde{A}_{\omega_j}\ltimes_{i\in\bar{\mathcal{N}}_{j}}x_i(t)\ltimes_{i\in\mathcal{N}^{c}_{j}}x_i(t). \\
		%~~~~=M_{\oplus_j}\textbf{K}_j(I_{2^{|\mathcal{N}_j|}}\otimes \bar{A}_j)\Phi_{2^{|\mathcal{N}_j|}}\ltimes_{i\in\bar{\mathcal{N}}_{j}}x_i(t)\ltimes_{i\in\mathcal{N}^{C}_{j}}x_i(t).\\
		%	u_j(t)=\textbf{K}_{j}\ltimes_{i\in\bar{\mathcal{N}}_{j}}x_i(t)\ltimes_{i\in\mathcal{N}^{C}_{j}}x_i(t).
		%\end{array}
		%	\right.
	\end{equation*}
	By considering Problem 2, one can find matrices $\tilde{A}_{\omega_j}$, $j\in\{1, \dots, \textbf{c}_1\}$ satisfying Eq. \eqref{pro-eq-constraint}. Further, one can obtain that 
	\begin{equation*}\label{proof-3}
		%	\left\{
		\begin{array}{rl}
			x_{\omega_j}(t+1)&=\hat{A}_{\omega_j}(I_{2^{|\bar{\mathcal{N}}_{\omega_j}|}}\otimes \textbf{1}_{2^{|\mathcal{N}_{\omega_j}^{c}|}}^\top)\ltimes_{i\in\bar{\mathcal{N}}_{j}}x_i(t)\ltimes_{i\in\mathcal{N}^{c}_{j}}x_i(t) \\
			&=\hat{A}_{\omega_j}\ltimes_{i\in\bar{\mathcal{N}}_{j}}x_i(t), j\in\{\omega_1, \dots, \omega_{\textbf{c}_1}\}. \\
		\end{array}
		%	\right.
	\end{equation*}
	This implies that in the interaction digraph $\textrm{G}_{id}=(\textrm{V}, \textrm{E})$, the NS-based distributed pinning controllers \eqref{pinning-controllers-step-1} will delete the edges from nodes $x_{\theta_1^{1}}, \cdots, x_{\theta_1^{\varepsilon_1}}$ to node $x_{\omega_1}$, $\cdots$, and the edges from nodes $x_{\theta_{\textbf{c}_1}^{1}}, \cdots, x_{\theta_{\textbf{c}_1}^{\varepsilon_{\textbf{c}_1}}}$ to node $x_{\omega_{\textbf{c}_1}}$. According to Problem 1, these deleted edges lead the interaction digraph $\textrm{G}_{id}=(\textrm{V}, \textrm{E})$ to be acyclic. Then, based on Lemma \ref{thm-graph}, system \eqref{pinning-controllers-step-1} is globally stabilized, which completes the proof. ~\IEEEQED

After considering Step 1, an NS-based distributed pinning control in the form of \eqref{pinning-controllers-step-1} is designed in order to guarantee global stabilization, but the designed pinning control \eqref{pinning-controllers-step-1} may not guarantee global stabilization to the given fixed point $\delta_{2^n}^\gamma$. Note that fixed points of BNs is used to represent stable configuration of cell types of genetic regulatory networks, such as cell death or unregulated growth. Thus, it is a significant issue to further design pinning control method to achieve global stabilization to a given fixed point, which will be studied in the following step. Under the NS-based distributed pinning control design in the form of (\ref{pinning-controllers-step-1}), with corresponding algebraic form (\ref{step-controlled}), one has the reduced system of \eqref{bn-alg}:
	\begin{equation}\label{deduced-eq}
		\left\{
		\begin{array}{l}
			x_{s}(t+1)=\hat{A}_{s}\ltimes_{i\in\bar{\mathcal{N}}_{s}}x_i(t), ~~s\in\{\omega_1, \dots, \omega_{\textbf{c}_1}\}, \\
			x_{t}(t+1)=A_{t}\ltimes_{i\in\mathcal{N}_{{t}}}x_i(t), ~t\in[1, n]\backslash\{\omega_1, \dots, \omega_{\textbf{c}_1}\}.
		\end{array}
		\right.
	\end{equation}
	As it is seen from \eqref{deduced-eq}, we have imposed state feedback pinning controllers on nodes $x_j$, $j\in\{\omega_1, \dots, \omega_{\textbf{c}_1}\}$, while there is no controller imposed on nodes $x_i$, $i\in[1, n]\backslash\{\omega_1, \dots, \omega_{\textbf{c}_1}\}$. Thus, the structure matrices for $x_i$, $i\in[1, n]\backslash\{\omega_1, \dots, \omega_{\textbf{c}_1}\}$, remain unchanged. However, in order to avoid confusion, we still denote $\hat{A}_i=A_i$, $\bar{\mathcal{N}}_i=\mathcal{N}_i$, $i\in[1, n]\backslash\{\omega_1, \dots, \omega_{\textbf{c}_1}\}$, for convenience of the following step.

\subsection{\textbf{Step 2: achieve global stabilization to a given state $\delta_{2^n}^\gamma$}} 
In the following, one needs to further design a state feedback pinning control to guarantee that \eqref{deduced-eq} will achieve global stabilization to the given fixed point $\delta_{2^n}^\gamma$. In order to further achieve global stabilization to the given fixed point $\delta_{2^n}^\gamma$, the following problem is considered to further determine the controlled nodes that are needed to achieve global stabilization to a given fixed point $\delta_{2^n}^\gamma$, the corresponding structure matrices of which will be changed under the design of state feedback pinning control.

	\textbf{Problem 3 (Determination of Pinning Nodes and Transformation of Structure Matrices):} Let $\delta_{2^n}^{\gamma}=\ltimes_{i=1}^{n}\delta_2^{\gamma_i}$. Find matrices $\check{A}_1\in\mathcal{L}_{2\times 2^{|\bar{\mathcal{N}}_1|}}, \dots, \check{A}_n\in\mathcal{L}_{2\times 2^{|\bar{\mathcal{N}}_n|}}$ and binary variables $\delta_1, \dots, \delta_n\in\mathcal{D}$ that minimize the cost function 
	\begin{equation}\label{cost-function}
		\textbf{c}_3\triangleq\sum_{i=1}^{n}\delta_i
	\end{equation}
	subject to the following conditions:
	\begin{equation}\label{cost-function-eq}
		%\left\{
		%\begin{array}{rl}
		%\delta_2^{\gamma_1}&=[(1-\delta_1)\hat{A}_1+\delta_1\check{A}_1]\ltimes_{j\in\bar{\mathcal{N}}_1}\delta_2^{\gamma_j}, \\
		%&\cdots \\
		%\delta_2^{\gamma_n}&=[(1-\delta_n)\hat{A}_n+\delta_1\check{A}_n]\ltimes_{j\in\bar{\mathcal{N}}_n}\delta_2^{\gamma_j}. \\
		%\end{array}
		%\right.
		\delta_2^{\gamma_i}=[(1-\delta_i)\times\hat{A}_i+\delta_i\times\check{A}_i]\ltimes_{j\in\bar{\mathcal{N}}_i}\delta_2^{\gamma_j}, ~~i\in[1,n].
	\end{equation}
	Problem 3 can be rewritten as an integer linear programming (ILP) problem, and can be solved by a suitable solver like Yices SMT Solver \cite{Kobayashi2016ITNNLSp1966}. Here, \eqref{cost-function-eq} is given to search structure matrices among $\hat{A}_{j}$ ($j\in[1, n]$) that does not satisfy the condition of the given fixed point $\delta_{2^n}^\gamma$ for \eqref{deduced-eq}. As one can see from \eqref{cost-function-eq}, if $\delta_1=1$, $\delta_2^{\gamma_1}=[(1-\delta_1)\hat{A}_1+\delta_1\check{A}_1]\ltimes_{j\in\bar{\mathcal{N}}_1}\delta_2^{\gamma_j}$ reduces to $\delta_2^{\gamma_1}=\check{A}_1\ltimes_{j\in\bar{\mathcal{N}}_1}\delta_2^{\gamma_j}$, this implies that node $x_1$ should be controlled and its structure matrix $\hat{A}_1$ should be changed to another structure matrix $\check{A}_1$ satisfying the condition of the given fixed point $\delta_{2^n}^\gamma$ for \eqref{deduced-eq}. If $\delta_1=0$, $\delta_2^{\gamma_1}=[(1-\delta_1)\hat{A}_1+\delta_1\check{A}_1]\ltimes_{j\in\bar{\mathcal{N}}_1}\delta_2^{\gamma_j}$ reduces to $\delta_2^{\gamma_1}=\hat{A}_1\ltimes_{j\in\bar{\mathcal{N}}_1}\delta_2^{\gamma_j}$, this implies that there is no need to transform the structure matrix $\hat{A}_1$ for node $x_1$. Thus, $\sum_{i=1}^{n}\delta_i$ represents the number of nodes requiring to be further controlled, and by minimizing the cost function $\textbf{c}_3=\sum_{i=1}^{n}\delta_i$, one can further determine the minimum number of controlled nodes. In addition, the determination of controlled nodes is given by searching non-zero variables among variables $\delta_1, \dots, \delta_n\in\mathcal{D}$.
	
	%
	%Here, an example is presented to illustrate Problem 1.
%	\begin{example}\label{example-1}
%		Here, a simple example is presented to illustrate Problem 1 and how to determine pining nodes for further achieving global stabilization to a given state $\delta_{2^n}^\gamma$. In order to improve the readability of the paper, the detailed illustrations can be found in Appendix A.~\IEEEQED
%		
%	\end{example}

A simple Example 3.1 is presented in Appendix A to illustrate problem 1 and how to determine pinning nodes for further achieving global stabilization to a given state$\delta_{2^n}^\gamma$.
	
	By considering Problem 3, one can find matrices $\check{A}_1\in\mathcal{L}_{2\times 2^{|\bar{\mathcal{N}}_1|}}, \dots, \check{A}_n\in\mathcal{L}_{2\times 2^{|\bar{\mathcal{N}}_n|}}$ and binary variables $\delta_1, \dots, \delta_n\in\mathcal{D}$, such that the cost function $\textbf{c}_3$ is minimized. Here, it is assumed that under the solution $\delta_{\tau_1}=\delta_{\tau_2}=\dots=\delta_{\tau_l}=1$, $\delta_{j}=0$ ($j\in\{1, \dots, n\}\backslash\{\tau_1, \dots, \tau_l\}$), the cost function $\textbf{c}_3$ is minimum, that is $\textbf{c}_3=l$, with the corresponding matrices $\check{A}_p$, $p\in\{\tau_1, \dots, \tau_l\}$. Since the determination of controlled nodes is given by non-zero variables among $\delta_1, \dots, \delta_n\in\mathcal{D}$, then nodes $x_p$, $p\in\{\tau_1, \dots, \tau_l\}$ will be further controlled in order to achieve global stabilization to the given fixed point $\delta_{2^n}^\gamma$.
	
	Then, further design an NS-based distributed pinning control to achieve global stabilization to a given fixed point $\delta_{2^n}^\gamma$. The controllers are imposed on nodes $x_p$, $p\in\{\tau_1, \dots, \tau_l\}$,
	\begin{equation}\label{-system-eq-22}
		\left\{
		\begin{array}{rll}
			x_p(t+1)&=&\hat{u}_p(t)\hat{\oplus}_p\hat{f}_p([x_j(t)]_{j\in\bar{\mathcal{N}}_p}), \\
			\hat{u}_p(t)&=&\hat{g}_{p}([x_j(t)]_{j\in\bar{\mathcal{N}}_p}),
		\end{array}
		\right.
	\end{equation}
	while the rest of the nodes $x_j$, $j\in\{1, \dots, n\}\backslash\{\tau_1, \dots, \tau_l\}$, remain unchanged. In addition, the structure matrices of logical functions $\hat{f}_p([x_j(t)]_{j\in\bar{\mathcal{N}}_p})$, $p\in\{\tau_1, \dots, \tau_l\}$, are $\hat{A}_{p}$, which are found in Problem 2. 
	
	Suppose that the structure matrices for logical functions $\hat{g}_p$, $p\in\{\tau_1, \dots, \tau_l\}$, are denoted by $\hat{K}_p\in\mathcal{L}_{2\times2^{|\bar{\mathcal{N}}_p|}}$, and the structure matrices of function $\hat{\oplus}_p$, $p\in\{\tau_1, \dots, \tau_l\}$, are denoted by $\hat{M}_{\hat{\oplus}_p}\in\mathcal{L}_{2\times4}$. Then, for $p\in\{\tau_1, \dots, \tau_l\}$, one can firstly obtain the algebraic form of (\ref{-system-eq-22}) as follows:
	\begin{equation}\label{-system-eq-23}
		\left\{
		\begin{array}{l}
			x_p(t+1)=\hat{M}_{\hat{\oplus}_p}\hat{u}_p(t)\hat{A}_p\ltimes_{j\in\bar{\mathcal{N}}_p}x_j(t), \\
			~~~~~~~~~~=\hat{M}_{\hat{\oplus}_p}\hat{K}_p(I_{2^{|\bar{\mathcal{N}}_p|}}\otimes \hat{A}_p)\Phi_{2^{|\bar{\mathcal{N}}_p|}}\ltimes_{j\in\bar{\mathcal{N}}_p}x_j(t), \\
			~~~~\hat{u}_p(t)=\hat{K}_{p}\ltimes_{j\in\bar{\mathcal{N}}_p}x_j(t).
		\end{array}
		\right.
	\end{equation}Thus, in order to determine matrices $\hat{K}_p\in\mathcal{L}_{2\times2^{|\bar{\mathcal{N}}_p|}}$, $\hat{M}_{\hat{\oplus}_p}\in\mathcal{L}_{2\times4}$, $p\in\{\tau_1, \dots, \tau_l\}$, one needs to solve the following equations for $p\in\{\tau_1, \dots, \tau_l\}$:
	\begin{equation}\label{step-controlled-system-eq-24}
		\check{A}_p=\hat{M}_{\hat{\oplus}_p}\hat{K}_p(I_{2^{|\bar{\mathcal{N}}_p|}}\otimes \hat{A}_p)\Phi_{2^{|\bar{\mathcal{N}}_p|}}.
	\end{equation}
	It is noted that Eq. \eqref{step-controlled-system-eq-24} is also solvable compared with \eqref{pro-eq-constraint-22}, which can be similarly proved using Theorem \ref{solvability}, thus it is implementable to design state feedback controllers to achieve global stabilization to the prescribed fixed point $\delta_{2^n}^\gamma$.

	In summary, for simplicity, here, we denote four sets $\mathcal{U}_{+}=\{\omega_1, \dots, \omega_{\textbf{c}_1}\}\bigcup\{\tau_1, \dots, \tau_l\}$, $\mathcal{U}_{-}=\{\omega_1, \dots, \omega_{\textbf{c}_1}\}\bigcap\{\tau_1, \dots, \tau_l\}$, $\mathcal{U}_{\omega}=\{\omega_1, \dots, \omega_{\textbf{c}_1}\}\backslash\mathcal{U}_{-}$, $\mathcal{U}_{\tau}=\{\tau_1, \dots, \tau_l\}\backslash\mathcal{U}_{-}$. Finally, after considering Steps 1 and 2, an NS-based distributed pinning control for global stabilization to a given fixed point $\delta_{2^n}^\gamma$ is designed as follows:
	\begin{equation}\label{finally-pinning-eq-1}
		\left\{
		\begin{array}{rl}
			x_{j_{1}}(t+1)&=\hat{u}_{j_{1}}\hat{\oplus}_{j_{1}}\left[u_{j_1}\oplus_{j_1}f_{j_1}([x_i(t)]_{i\in\mathcal{N}_{j_1}})\right], j_1\in\mathcal{U}_{-}, \\
			x_{j_{2}}(t+1)&=\left[u_{j_1}\oplus_{j_2}f_{j_2}([x_i(t)]_{i\in\mathcal{N}_{j_2}})\right], ~~~~~~~~ j_2\in\mathcal{U}_{\omega}, \\
			x_{j_{3}}(t+1)&=\left[\hat{u}_{j_3}\hat{\oplus}_{j_3}f_{j_3}([x_i(t)]_{i\in\mathcal{N}_{j_3}})\right], ~~~~~~~~~ j_3\in\mathcal{U}_{\tau}, \\
			x_{j_{4}}(t+1)&=f_{j_4}([x_i(t)]_{i\in\mathcal{N}_{j_4}}), ~~~~~~~~~~~~~j_4\in[1, n]\backslash\mathcal{U}_{+}. \\
		\end{array}	
		\right.
	\end{equation}
	Here, $u_j$ and $\oplus_j$, $j\in\{\omega_1, \dots, \omega_{\textbf{c}_1}\}$, are controllers obtained in Step 1, while $\hat{u}_j$ and $\hat{\oplus}_j$, $j\in\{\tau_1, \dots, \tau_l\}$, are controllers obtained in Step 2. Hence, this completes the main procedures of designing an NS-based distributed pinning control for global stabilization to a given fixed point $\delta_{2^n}^\gamma$.

	\begin{remark}
		During the past few decades, many fundamental results using the state transition matrix $L$ have been established \cite{Laschov2012Ap1218,chen2019ITCNSp1379,wu2017ITCNSp1100,Meng2017ITNNLSp}. Unfortunately, there still exists several disadvantages concerning the state transition matrix $L$. One is that using system $x(t+1)=Lx(t)$, the information of network structure is missing, such as nodes' connection, while matrix $L$ only reflects the state transition digraph. Another one is that the dimension of $L$ grows drastically with the size of network, which implies that the existing methods are difficult to be implemented for large-scale BNs. In this paper, a new approach for designing an NS-based distributed pinning control approach is proposed based on network structure of BNs without using the state transition matrix $L$. This paper aims to solve the problems of computational complexity (reduced from $O(2^n\times 2^n)$ to $O(2\times 2^K)$, where $n$ is the number of nodes and $K$ is the largest number of in-neighbors of nodes). 
	\end{remark}

	\section{Applications to T-LGL survival signaling and T-Cell receptor signaling networks}\label{simulations}
	
	In this section, several biological networks are presented to demonstrate the validity of the obtained results, the first example is a network with 6 nodes, the second one has 90 nodes, and the last one can be found in Appendix.

	\begin{example}[\textbf{T-LGL Survival Signaling Network}]\label{sim-exam}
		Consider a reduced network in the T-LGL survival signaling network, which consists of six nodes, that is S1P, FLIP, Fas, Ceramide, DISC, Apoptosis \cite{Campbell2014BSSp53}. Here, $x_1, x_2, x_3, x_4, x_5, x_6$ are used to represent these six nodes, respectively. 
		%	The interaction digraph $\textrm{G}_{id}=(\textrm{V}, \textrm{E})$ is shown in Fig. \ref{sim-2--2-1} (a). 
		The logical dynamics are given as follows:
		\begin{equation}\label{sim-1}
			\left\{
			\begin{array}{l}
				x_1({\ast})=\neg (x_4\vee x_6), ~~~\quad x_2({\ast})=\neg (x_5\vee x_6), \\
				x_3({\ast})=\neg (x_1\vee x_6), ~~~\quad x_4({\ast})=x_3\vee \neg (x_1\vee x_6),\\
				x_5({\ast})=[x_4\vee(x_3\wedge \neg x_2)]\wedge \neg x_6, ~~x_6({\ast})=x_5\vee x_6.\\
			\end{array}
			\right.
		\end{equation}
		The symbol ($\ast$) indicates the state of next step of the marked node to reduce space. For each node $x_i$, $i\in[1, 6]$, one has its corresponding in-neighbors, that is $\mathcal{N}_1=\{4, 6\}$, $\mathcal{N}_2=\{5, 6\}$, $\mathcal{N}_3=\{1, 6\}$, $\mathcal{N}_4=\{1, 3, 6\}$, $\mathcal{N}_5=\{2, 3, 4, 6\}$ and $\mathcal{N}_6=\{5, 6\}$. By simple calculations, one can conclude that system \eqref{sim-1} is not globally stable to state $\delta_{64}^{31}$.  Now, we consider how to design an efficient NS-based distributed pinning control to achieve global stabilization to state $\delta_{64}^{31}$.
		
		Thus, by considering Steps 1 and 2, one can finally design the following NS-based distributed pinning controllers $u_1, u_5, u_6, \hat{u}_1$, which are imposed on nodes $x_1, x_5$ and $x_6$, 
		\begin{equation}\label{controllers-1-5-6-98}
			\left\{
			\begin{array}{l}
				x_1({\ast})=\hat{u}_1\wedge \neg[u_1\leftrightarrow \neg (x_4\vee x_6)], \\
				x_2({\ast})=\neg (x_5\vee x_6), ~~~~~~~~~x_3({\ast})=\neg (x_1\vee x_6), \\
				x_4({\ast})=x_3\vee \neg (x_1\vee x_6), ~~~~x_6({\ast})=u_6\vee x_5\vee x_6,\\
				x_5({\ast})=u_5\leftrightarrow \{[x_4\vee(x_3\wedge \neg x_2)]\wedge \neg x_6\}. \\
			\end{array}
			\right.
		\end{equation} 
		The corresponding NS-based distributed pinning controllers $u_1, u_5, u_6, \hat{u}_1$ are designed as follows:
		\begin{equation}\label{controllers-1-5-6-lo-97}
			\left\{
			\begin{array}{l}
				u_1=x_4\rightarrow x_6, ~~\hat{u}_1=x_6, ~~u_6=\neg x_5\wedge \neg x_6, \\
				%\end{array}
				%\begin{array}{l}
				u_5=[x_2\wedge (x_3\vee x_6)]\vee \neg x_2.\\
			\end{array}
			\right.
		\end{equation} 
		Due to the page limitation, the detailed derivations for the above pinning control \eqref{controllers-1-5-6-98} and \eqref{controllers-1-5-6-lo-97} are omitted here, which can be obtained using the proposed method in this paper. Thus, by controlling nodes $x_1, x_5, x_6$ and four corresponding NS-based distributed pinning controllers $u_1, u_5, u_6, \hat{u}_1$ in the form of \eqref{controllers-1-5-6-98} and \eqref{controllers-1-5-6-lo-97}, system \eqref{controllers-1-5-6-98} is globally stabilized to state $\delta_{64}^{31}$. The state transition graphs for both \eqref{sim-1} and \eqref{controllers-1-5-6-98} under pinning control \eqref{controllers-1-5-6-lo-97}, are shown in Fig. \ref{6nodes_stategraph}(a) and Fig. \ref{6nodes_stategraph}(b). Using the traditional $L$-based pinning control method \cite{LiF2015IToNNaLSp1585,LiF2016IToCaSIEBp309,LuJ2016IToACp1658,Li2017ITNNLS,Li2017ITCYT,li2020ITCNSp201}, that is to change certain columns of matrix $L$, one needs at least controlling nodes $x_1, x_2, x_5, x_6$. However, using the proposed NS-based distributed pinning control method in this paper, one can achieve global stabilization by controlling nodes $x_1, x_5, x_6$. Without using matrix $L$, the computational complexity can be reduced, and the NS-based distributed pinning control method also leads to a lower dimensional controller design than the traditional $L$-based pinning control design. 
		\begin{figure}\centering
			\includegraphics[width=8.6cm,height=4cm]{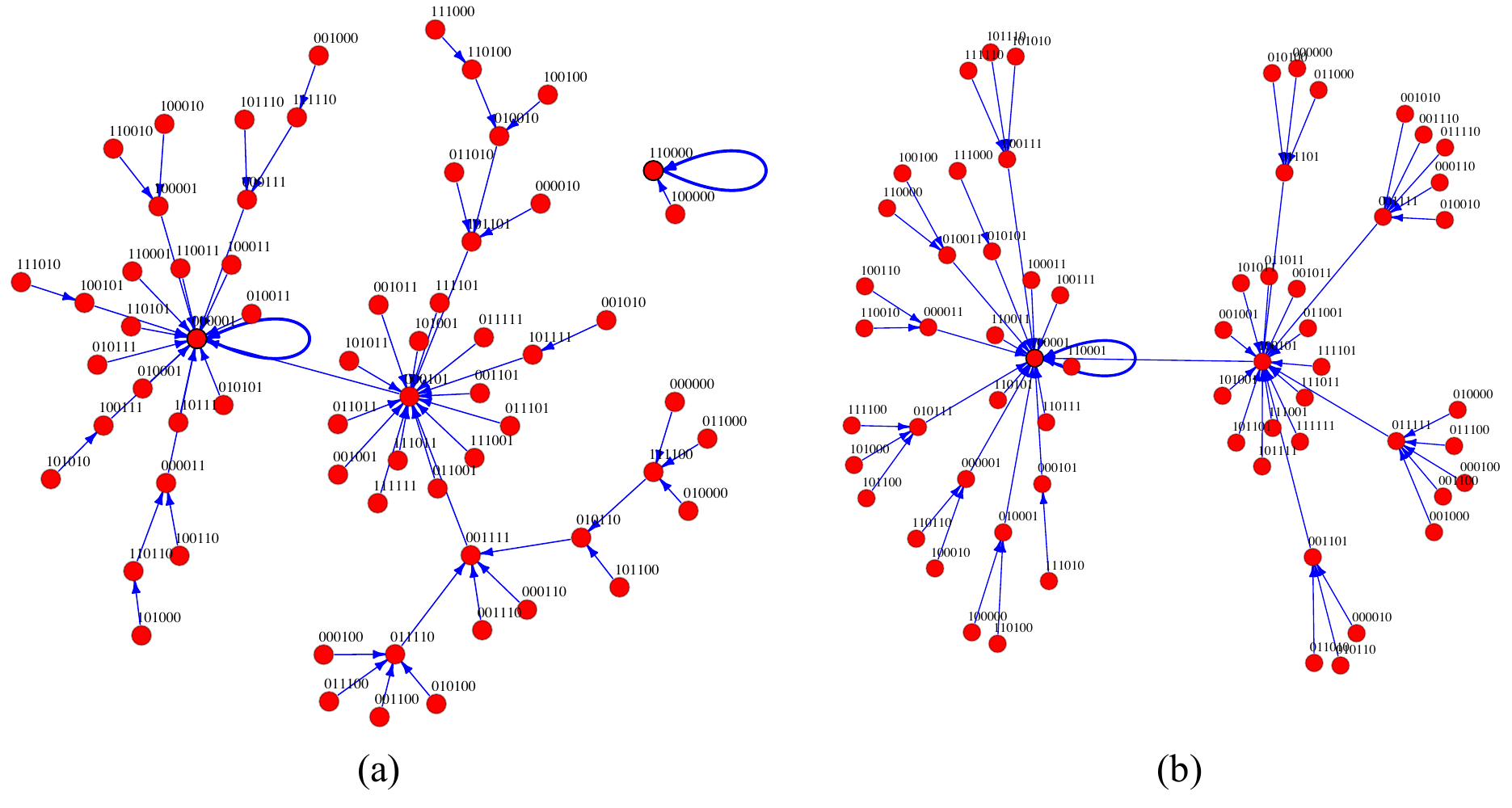}
			\caption{(a) State transition digraph of \eqref{sim-1} with two fixed points ($(0,0,0,0,0,1)$ and $(1,1,0,0,0,0)$); (b) State transition graph of \eqref{controllers-1-5-6-98} under pinning control strategies \eqref{controllers-1-5-6-lo-97} with a unique fixed point $(1, 0, 0, 0, 0, 1)$. Each node represents a state of the network, and each arrow is a state transition.}\label{6nodes_stategraph}
		\end{figure}
		%	\begin{figure}\centering
		%		\includegraphics[width=8cm,height=3.3cm]{Rplot-6nodes_pin.pdf}
		%		\caption{The state transition graph of system \eqref{controllers-1-5-6-98} under NS-based distributed pinning control strategies \eqref{controllers-1-5-6-lo-97}, which has one unique attractor (the unique fixed point $(1, 0, 0, 0, 0, 1)$). Each node represents a state of the network, and each arrow is a state transition. }\label{6nodes_stategraph_pin}
		%	\end{figure}

	\end{example}

\begin{example}[\textbf{T-Cell Receptor Signaling Network}]
Consider a large Boolean model of a cellular network in \cite{zou2013algorithm}, which well model the logical dynamics of T-Cell receptor signaling model. In fact, the proposed NS-based distributed pinning control method has been tested in a network model consisting of 90 nodes denoted by $x_1, \dots, x_{90}$ \cite{zou2013algorithm}, the detailed logical functions can be found in Table 3 in \cite{zou2013algorithm} and the interaction digraph is shown in Fig. \ref{90nodes-figure}(a). It is said in \cite{Saez2007PCBpe163}, ``\emph{It is, to the best of our knowledge, the largest Boolean model of a cellular network to date}''. Due to the page limitation, the corresponding logical functions are omitted here, we refer the readers to \cite{Saez2007PCBpe163}. According to the proposed NS-based distributed pinning control, design an NS-based distributed pinning control just on 15 nodes $x_1$, $x_2, x_3$, $x_4, x_9$, $x_{12}, x_{21}$, $x_{22}, x_{38}$, $x_{47}, x_{52}$, $x_{68}, x_{69}, x_{78}, x_{79}$. The corresponding distributed pinning control for these 15 nodes are given below: $x_1(\ast)=u_1\vee x_1, x_2(\ast)=u_2\vee x_2, x_3(\ast)=u_3\vee x_3, x_4(\ast)=u_4\wedge x_4$, $x_9(\ast)=u_9\vee [x_2\wedge x_4\wedge \neg x_6\wedge \neg x_7]$, $x_{12}(\ast)=u_{12}\vee x_{17}, x_{21}(\ast)=u_{21}\wedge x_{21}$, $x_{22}(\ast)=u_{22}\vee x_{22}, x_{38}(\ast)=u_{38}\vee x_{38}$, $x_{47}(\ast)=u_{47}\vee x_{47}, x_{52}(\ast)=u_{52}\vee x_{49}$, $x_{68}(\ast)=u_{68}\vee x_{68}, x_{69}(\ast)=u_{69}\wedge x_{69}$, $x_{78}(\ast)=u_{78}\vee x_{78}, x_{79}(\ast)=u_{79}\vee x_{79}$; and the logical dynamics for each controller are given below: $u_1=\neg x_1, u_2=\neg x_2, u_3=\neg x_3, u_4(\ast)=\neg x_4$, $u_9=x_2\wedge x_4\wedge \neg x_6$, $u_{12}=\neg x_{17}, u_{21}=\neg x_{21}$, $u_{22}=\neg x_{22}, u_{38}=\neg x_{38}$, $u_{47}=\neg x_{47}, u_{52}=\neg x_{49}$, $u_{68}=\neg x_{68}, u_{69}=\neg x_{69}$, $u_{78}=\neg x_{78}, u_{79}=\neg x_{79}$. Fig. \ref{90nodes-figure}(b) shows part of the state transition digraph randomly from 25 initial states before pinning control, while Fig. \ref{90nodes-figure}(c) shows part of the state transition digraph randomly from 500 initial states after control. This implies that the proposed NS-based distributed pinning control in this paper is effective in certain biological networks, while it is almost impossible using traditional $L$-based pinning control \cite{LiF2015IToNNaLSp1585,LiF2016IToCaSIEBp309,LuJ2016IToACp1658,Li2017ITNNLS,Li2017ITCYT,li2020ITCNSp201}. This is because using the traditional $L$-based pinning control, one needs to calculate the state transition matrix $L$ with dimension $2^{90}\times 2^{90}$ in order to analyze and design pinning control strategy, which leads to an extremely high computational complexity.   
\begin{figure}\centering
	\includegraphics[width=8.6cm,height=4cm]{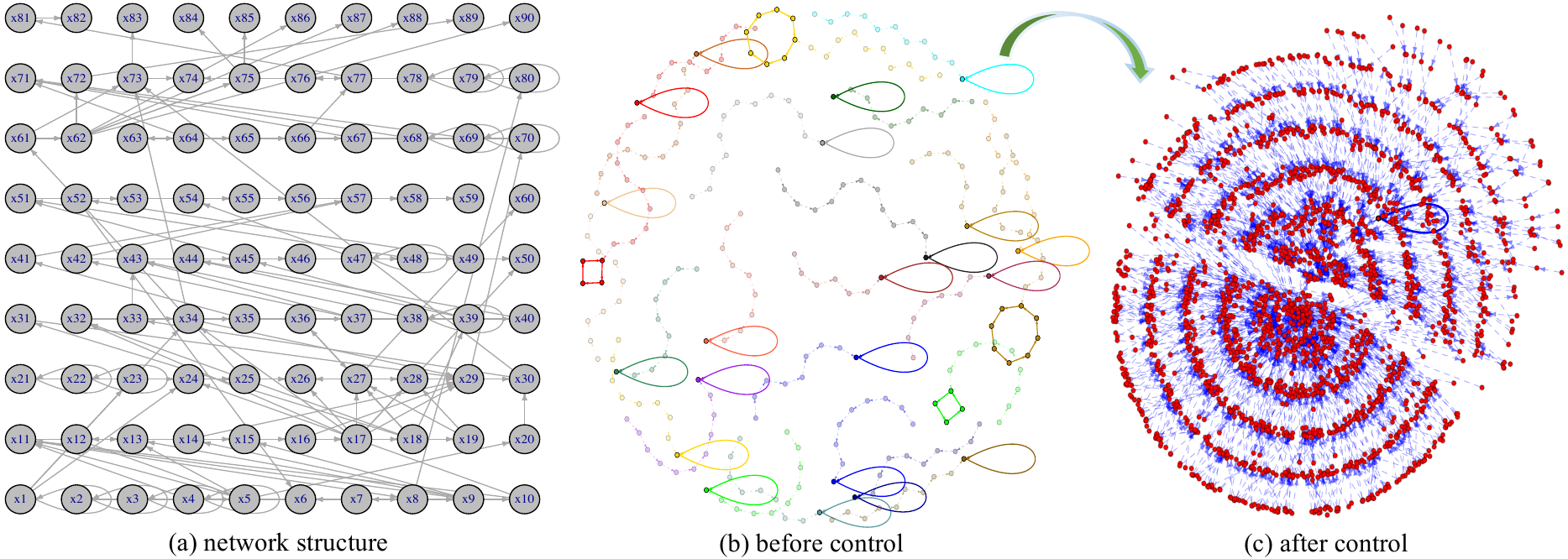}
	\caption{(a) Network structure of T-Cell receptor signaling Boolean model \cite{zou2013algorithm}; (b) Part of state transition digraph randomly from 25 initial states before pinning control; (c) Part of state transition digraph randomly from 500 initial states after pinning control. }\label{90nodes-figure}
\end{figure}	
\end{example}

\begin{example}
Another two biological networks, a reduced network of T-LGL signaling network with 18 nodes \cite{Saadatpour2011PCBpe1002267} and a network model of survival signaling in large granular lymphocyte leukemia with 29 nodes \cite{Zhang2008PNASp16308}, have been tested using the proposed NS-based distributed pinning control method in this paper, which can be found in Appendix B. We refer the readers to \cite{Saadatpour2011PCBpe1002267,Zhang2008PNASp16308} for the detailed logical dynamics.
\end{example}

\begin{remark}
	It has been formulated that gene regulatory networks in most biological systems are sparse \cite{Jeong2013Np41}, which implies that it is practical and reasonable in determining certain feasible feedback arc sets over the interaction digraph of BNs. Just as shown in \cite{Thieffry1998Bp433}, the mean connectivity in Escherichia coli is found between 2 and 3, which shows a rather loosely interconnected structure.
	Thus, the proposed edge-deleting methodology is practical and reasonable in the majority of gene networks since generally they are sparse. In addition, the proposed edge-deleting methodology can model the action of a drug that inactivates the corresponding interaction among two gene products \cite{Murrugarra2016BSBp94}, while deleting nodes of BNs can model the control action blocking of effects of products of genes associated to these nodes. 
\end{remark}

	\section{Conclusion}\label{conclusion}
	%%%%%%%%%%%%%%%%%%%%%%%%%%%%%%%%%%%%%%%%%%%%%%%%%%%%%%%%%%%%%%%%%%%%%%%%%%%%%%%%%%%%%%%%%%%%%%%%%
	
	In this paper, a new approach for designing NS-based distributed pinning control has been presented to achieve global stabilization to any given fixed point, without using state transition matrix. The NS-based distributed pinning control is designed based on local neighbors of controlled nodes. In addition, the computational complexity is dramatically reduced, which can be efficiently applied to large-scale BNs, such as T-Cell receptor signaling Boolean model with 90 nodes. 
	A new perspective has been proposed combining the network structure with NS-based distributed pinning control design, which aims to solve the problem of computational complexity for large-scale networks. 
	
	To sum up, by comparing the simulations on certain BNs with different sizes, the traditional $L$-based pinning control \cite{LiF2015IToNNaLSp1585,LiF2016IToCaSIEBp309,LuJ2016IToACp1658,Li2017ITNNLS,Li2017ITCYT,li2020ITCNSp201} has limitations on the following certain aspects:
	\begin{enumerate}
		\item[I.] Calculations of algebraic system $x(t+1)=Lx(t)$ is always needed, where the state transition matrix $L$ has a dimension $2^n\times 2^n$. Thus, when the number of network nodes $n$ becomes larger, the computational complexity ($O(2^n\times 2^n)$) will be more complicate;
		\item[II.] Since the traditional $L$-based pinning control is designed based on the transformation of matrix $L$, then $L$-based pinning controllers depends on all the nodes and also are in the form of $u(t)=g(x_1(t), \dots, x_n(t))$. Thus, when network nodes $n$ become larger, $L$-based pinning controllers will be more complicated and have higher dimensions;
		\item[III.] $L$-based pinning controlled nodes are determined by matrix $L$, when the dimension of $L$ becomes larger, pinning controlled nodes will be difficult to be found. 
	\end{enumerate}
	The above limitations are well solved by the proposed NS-based distributed pinning control, which reduces the high computation complexity to $O(2\times 2^K)$ ($K$ is the largest number of neighbors among each node). By observing simulations on several biological networks, the proposed approach is effective in biological networks with sparse neighbors. In addition, the proposed pinning control technique can be applied to reduce the computational complexity of controllability and observability of BNs and BCNs.

%	\begin{enumerate}
%		\item[I.] In this paper, during the procedures of designing NS-based distributed pinning control, the calculations of $x(t+1)=Lx(t)$ is avoided, which reduces the high computation complexity $O(2^n\times 2^n)$ from $O(2\times 2^K)$ ($K$ is the largest number of neighbors among each node); 
%		\item[II.] Using the feedback arc set in the network structure of BNs, pinning controlled nodes can be easily determined under the information of network structure; 
%		\item[III.] By using neighbors' local information not on overall nods' information, NS-based distributed pinning controllers only depends on local neighbors and are in the form of $u(t)=g(x_{i_1}(t), x_{i_2}(t), \dots, x_{i_p}(t))$, $\{i_1, i_2, \dots, i_p\}\subseteq\{1, 2, \dots, n\}$ are the neighbors of controlled nodes. Thus, compared with $L$-based pinning controllers \cite{LiF2015IToNNaLSp1585,LiF2016IToCaSIEBp309,LuJ2016IToACp1658,Li2017ITNNLS,Li2017ITCYT,li2020ITCNSp201}, NS-based distributed pinning controllers are much simpler and easier to be designed, and also have lower dimensions.
%	\end{enumerate}
	
\section*{Appendix}

\subsection{Example 3.1}
Consider a BN with 3 nodes, and its equivalent algebraic form is given below, 
\begin{equation}\label{example-1-eq-2}
%	\left\{
%	\begin{array}{l}
x_1(t+1)=\hat{A}_1x_3(t), 
x_2(t+1)=\hat{A}_2x_3(t), 
x_3(t+1)=\delta_2^1.
%	\end{array}
%	\right.
\end{equation}
Here, $\hat{A}_1=\delta_2[2, 1]$, $\hat{A}_2=\delta_2[1, 2]$. Let $x(t)=\ltimes_{j=1}^3x_j(t)$, one has that
$x(t+1)=Lx(t)$,
where $L=\delta_8[5,3,5,3,5,3,5,3]$. According to Lemma \ref{thm-graph}, \eqref{example-1-eq-2} is globally stable to state $\delta_{8}^5$. Suppose that state $\delta_{8}^5$ is not the given fixed point we want while state $\delta_8^7$ is, here we consider how to achieve global stabilization of \eqref{example-1-eq-2} to state $\delta_8^7$ by transforming certain structure matrices of \eqref{example-1-eq-2}. Here, we consider Problem 3 and the constraint conditions \eqref{cost-function-eq} to find matrices $\check{A}_1\triangleq\left(\begin{array}{cc}a_{11}&a_{12} \\1-a_{11}&1-a_{12} \end{array}\right)\in\mathcal{L}_{2\times 2}, \check{A}_2\triangleq\left(\begin{array}{cc}b_{11}&b_{12} \\1-b_{21}&1-b_{22} \end{array}\right)\in\mathcal{L}_{2\times 2}, \check{A}_3\triangleq\left(\begin{array}{c}c_{11}\\1-c_{11} \end{array}\right)\in\mathcal{L}_{2\times 1}$ and binary variables $\delta_1, \delta_2, \delta_3\in\mathcal{D}$, such that $\textbf{c}_3\triangleq\sum_{i=1}^{3}\delta_i$ is minimum. Thus, the constraint conditions \eqref{cost-function-eq} can be simplified as follows:
\begin{equation}\label{constraint-conditions-eq-2}
%		\left\{
%		\begin{array}{l}
\delta_1a_{11}=0, ~
1-\delta_2+\delta_2b_{11}=0,~ 
1-\delta_3+\delta_3c_{11}=0. 
%		\end{array}\right.
\end{equation}		
Then, Problem 3 is converted to find the indexes $a_{11}, a_{12}, b_{11}, b_{12}, c_{11}\in\mathcal{D}$, and binary variables $\delta_1, \delta_2, \delta_3\in\mathcal{D}$, minimizing the cost function 
$
\textbf{c}_3=\sum_{i=1}^{3}\delta_i
$
subject to \eqref{constraint-conditions-eq-2}. Thus, one can find the minimum cost $\textbf{c}_3=1$ under the possible solution $\delta_1=\delta_3=0$, $\delta_2=1$, and logical matrix $\check{A}_2$ satisfying $\textrm{Col}_{1}(\check{A}_2)=\delta_{2}^2$.
For example, there exists a feasible solution for $\check{A}_2$ as $\check{A}_2=\delta_2[2, 1]$. In fact, system \eqref{example-1-eq-2} could be globally stabilizable to any state by transforming structure matrices according to Problem 3. For example, consider global stabilization of \eqref{example-1-eq-2} to state $\delta_8^3$, one can also find $\delta_1=\delta_2=1$, $\delta_2=0$ such that $\textbf{c}_3=\sum_{i=1}^{3}\delta_i$ equals to 2, and a feasible solution is matrices $\check{A}_1=\delta_2[1, 2]$ and $\check{A}_2=\delta_2[2, 1]$.

\subsection{Example 4.3 (Further Applications)}
A brief report is given here on applying the proposed method to reduced T-LGL signaling network with 18 nodes \cite{Saadatpour2011PCBpe1002267} and large granular lymphocyte leukemia with 29 nodes \cite{Zhang2008PNASp16308}. Consider the reduced network of T-LGL signaling network with 18 nodes, CTLA4, TCR, CREB, IFNG, P2, GPCR, SMAD, Fas, aFas, Cermide, DISC, Caspase, FLIP, BID, IAP, MCL1, S1P, Apoptosis \cite{Saadatpour2011PCBpe1002267}. Using the proposed NS-based distributed pinning control method in this paper, an NS-based distributed pinning controllers is designed to achieve global stabilization on nodes CTLA4, P2, DISC, S1P and Apoptosis: $\textrm{CTLA4}=u_1\vee \textrm{TCR}, 
\textrm{P2}=u_2\wedge (\textrm{IFNG}\vee \textrm{P2}), 
\textrm{DISC}=u_3\vee  [\textrm{Cermide}\vee (\textrm{Fas}\wedge \neg \textrm{FLIP})], 
\textrm{S1P}=u_4\vee \neg \textrm{Cermide}, 
\textrm{Apoptosis}= u_5\wedge  (\textrm{Caspase} \vee \textrm{Apoptosis})$.
In addition, the corresponding NS-based distributed pinning controllers are designed below: $u_1=\neg \textrm{TCR}, 
u_2=\neg (\textrm{IFNG}\vee \textrm{P2}), 
u_3=\textrm{Fas} \vee (\neg \textrm{Fas} \wedge \textrm{Cermide}), 
u_4=\textrm{Cermide}, 
u_5=\textrm{Caspase}$.

As for the network model of survival signaling in large granular lymphocyte leukemia with 29 nodes (IL15, RAS, ERK, JAK, IL2RBT, STAT3, IFNGT, FasL, PDGF, PDGFR, PI3K, IL2, BcIxL, TPL2, SPHK, S1P, sFas, Fas, DISC, Caspase, Apoptosis, LCK, MEK, GZMB, IL2RAT, FasT, RANTES, A20, FLIP) \cite{Zhang2008PNASp16308}, an NS-based distributed pinning controllers is designed on nodes IL15, PDGF, SPHK, in order to achieve global stabilization: $\textrm{IL15}=u_1\wedge \textrm{IL15}, \textrm{PDGF}=u_2\vee \textrm{PDGF}, \textrm{SPHK}=u_3\wedge (\textrm{PI3K}\wedge \textrm{S1P})$.
The logical dynamics of NS-based distributed pinning controllers $u_1, u_2, u_3$, are $u_1=\neg \textrm{IL15}, u_2=\neg \textrm{PDGF}, u_3=\textrm{PI3K}$.

\bibliographystyle{ieeetr}
\bibliography{survey2}
\end{document}